\documentclass{bmcart}

\usepackage{amsthm,amsmath,cite}
\usepackage[utf8]{inputenc} 

\usepackage{multirow}   
\usepackage{subcaption} 
\usepackage{colortbl}   
\usepackage{graphicx}   

\renewcommand{\th}{\text{\scriptsize th}}

\startlocaldefs
\endlocaldefs
\graphicspath{{Figures/}}

\begin{document}

\begin{frontmatter}

\begin{fmbox}
\dochead{Regular Article}


\title{Are the different layers of a social network conveying the same information?}

\author[
   addressref={aff1}                   
]{\inits{AM}\fnm{Ajaykumar} \snm{Manivannan}}
\author[
   addressref={aff1}
]{\inits{WQY}\fnm{W. Quin} \snm{Yow}}
\author[
   addressref={aff1},
   noteref={n1},                        
   email={bouffanais@sutd.edu.sg}
]{\inits{RB}\fnm{Roland} \snm{Bouffanais}}
\author[
   addressref={aff2,aff3},
   corref={aff1,aff2,aff3},                    
   noteref={n1},                        
   email={alain.barrat@cpt.univ-mrs.fr}
]{\inits{AB}\fnm{Alain} \snm{Barrat}}


\address[id=aff1]{%
  \orgname{Singapore University of Technology and Design}, 
 \street{8 Somapah Road},      
 \postcode{487372},                 
  \cny{Singapore}                    
}
\address[id=aff2]{%
  \orgname{Aix Marseille Univ, Universit\'e de Toulon, CNRS, CPT},
  \city{Marseille},
  \cny{France}
}
\address[id=aff3]{%
  \orgname{Data Science Laboratory, Institute for Scientific Interchange (ISI) Foundation},
  \city{Torino},
  \cny{Italy}
}

\begin{artnotes}
\note[id=n1]{Equal contributor} 
\end{artnotes}

\end{fmbox}


\begin{abstractbox}

\begin{abstract} 
Comprehensive and quantitative investigations of social theories
and phenomena increasingly benefit from the vast breadth of data
describing human social relations, which is now available within the realm of computational social science. Such data are, however, typically
proxies for one of the many interaction layers composing
social networks, which can be defined
in many ways and are typically composed of 
communication of various types (e.g., phone calls, face-to-face communication, etc.). As a result, 
many studies focus on one single layer, corresponding to the 
data at hand. Several studies have, however, shown that these layers
are not interchangeable, despite the presence of a certain level of correlations between them.
Here, we investigate whether different layers of interactions
among individuals lead to similar conclusions with respect to
the presence of homophily patterns in a population---homophily
represents one of the widest studied phenomenon in social networks.
To this aim, we consider a dataset describing interactions and links of
various nature in a population of Asian students with 
diverse nationalities, first language and gender. We study 
homophily patterns, as well as their temporal evolutions in each layer of the
social network.
To facilitate our analysis, we put forward a general method
to assess whether the homophily patterns observed in one layer inform us about patterns in another layer. For instance, our study reveals that three network layers---cell phone communications, questionnaires about friendship, and trust 
relations---lead to similar and consistent results despite some minor discrepancies. The homophily patterns of the co-presence network layer, however, does not yield any meaningful information about other network layers.

\end{abstract}


\begin{keyword}
\kwd{social networks}
\kwd{multilayer networks}
\kwd{temporal homophily}
\end{keyword}


\end{abstractbox}
%

\end{frontmatter}



\section*{Introduction}

Mining and analyzing social networks in various contexts 
yield important insights towards a better fundamental knowledge and understanding of human behavior~\cite{Wasserman:1994}. Data on social networks have allowed researchers to investigate social theories and effects such as homophily, influence,
triadic closure, etc. Data also help design data-driven models of human interactions, which can be used to describe the many processes taking place in a given population, such as information spreading, coordination, consensus formation, or spread of infectious diseases~\cite{Barrat:2008}.
Accurate descriptions of social interactions are therefore crucial to shed light on the most relevant mechanisms at work in these processes, and for instance
to understand the factors determining if a rumor will spread, or what are the best measures to contain the spread of a disease.

Within a given population, however, several networks of social interactions can be defined: e.g., friendship relations, patterns
of communications, co-presence, face-to-face interactions. These different types of relations form a multilayer network~\cite{DeChoudhury:2010,Mollgaard:2016}, for which each layer can be explored using possibly different methods. Friendship relations are typically mined through surveys,
physical interactions and proximity by diaries or more recently using wearable sensors~\cite{Read:2012,Barrat:2015}, and communication patterns are extracted from mobile phone call records~\cite{Onnela:2007,Miritello:2013,Saramaki:2014}. In recent times in particular, technological developments have allowed researchers
to gather increasing amounts of digital data on face-to-face contacts, phone communication patterns and online relationships, at widely different scales in terms of population size,
space and time resolution. These data 
have been widely used to investigate the structure of social networks, the patterns of social interactions and social theories, such as
the strength of weak ties~\cite{Onnela:2007}, homophily patterns (the tendency of individuals to have social links with similar individuals, with respect to gender, nationality, social class, etc.~\cite{McPherson:2001}) 
\cite{Kossinets:2009,Aiello:2010,Stehle:2013,Palchykov:2012,Kovanen:2013,Jo:2014},
mechanisms of link formation and persistence~\cite{Kossinets:2009,Aiello:2010,Navarro:2017}, social
strategies linked to limited attention capacity~\cite{Miritello:2013}, etc.

In most cases, such studies are based on data describing one specific layer of the multilayer network characterizing social interactions, and consider this layer as a proxy of
``the" social network of the population under study. Several research groups have, however, managed in recent years to access simultaneously more than one layer of interactions
in various population groups, showing in each case that these layers are correlated but not equivalent 
\cite{Stopczynski:2014,Sekara:2014,Mastrandrea:2015,Leecaster:2016,Smieszek:2016,Mollgaard:2016,Boonstra:2017,Mones:2017}. For instance,
a comparison between face-to-face contacts measured by sensors and friendship relations obtained through surveys has shown that
the distribution of contact durations are broad both for pairs of friends and pairs of non-friends, even if the longest contacts occur between friends~\cite{Mastrandrea:2015}. In addition, a comparison between proximity events and online social links has shown that a simple thresholding procedure
retaining only the strongest proximity links is not enough to determine online friendship~\cite{Sekara:2014}. Furthermore, a recent study of communication, online links,
and proximity events has highlighted that these layers differ and cannot be reduced to a single channel of interaction~\cite{Mones:2017}. 

Despite these well-accepted and known differences between the ``social networks" defined through different proxies \cite{DeChoudhury:2010}, 
many authors have argued that close relationships correspond to both higher frequencies of face-to-face contacts and phone communication
\cite{Hill:2003,Roberts:2009,Kossinets:2009,Palchykov:2012,Jo:2014,Saramaki:2014}. It is, for instance, often assumed that the most important
relationship of an individual can be captured by his/her mobile phone records, and that the ``best friend" of an individual is the person he or she is in most contact with.
Some evidence to support this assumption has come from surveys \cite{Hill:2003,Roberts:2010} or from comparison between surveys and mobile
phone records \cite{Saramaki:2014}, which are, however, rarely available for the same population.

It is thus important to gather and investigate datasets containing multiple layers of social interactions, to better ground
such assumption and assess the extent of its validity. It is worth highlighting that the number of datasets offering multiple layers of interactions, enriched with metadata describing individual characteristics, remains extremely limited. Moreover, it is crucial to investigate whether, given that the layers of interactions
are correlated but not equivalent, socially relevant patterns and theories can be reliably assessed from one layer only. If it is indeed
the case, then for a given population the data that is most conveniently accessible or that offers the best resolution can safely be used to explore such issues.
Here, we focus on homophily along a range of individual characteristics, as one of the most explored patterns structuring social networks~\cite{McPherson:2001}.
A recent study has shown some notable differences in the strength of homophilous patterns in different communication channels in a population of European students~\cite{Mollgaard:2016}. We investigate this particular issue in a diverse population of Asian students of various
nationalities in a university of Singapore,
for which we have access to phone communication records, co-presence events, and friendship and trust relations over one full calendar year.
Detailed metadata about gender, nationality, first spoken language, academic performance and psychological traits are also available, allowing us
to assess homophily and its temporal evolution along multiple traits and multiple layers of social relationships. 
We put forward a methodology to 
systematically compare homophily patterns across layers,
as observed through different indicators and with respect to different attributes, and apply this methodology to our dataset. In this case, we show that patterns of homophily in 
the co-presence layer do not inform us on the patterns
in other layers, while the patterns observed in the
communication network and in the networks of 
friendship and trust obtained from surveys, although not 
equal, are informative of each other.

\section*{Data and methods}\label{sec:methods}

We consider data collected in a Singapore university during one full academic year---three consecutive terms separated by short breaks---and concerning two different cohort classes 
studying in the same campus and staying at the same on-campus hostel. The data consists in several types of relationship between students, as well as in metadata about each student.

First, the $35$ participating students used 
Android smartphones preinstalled with a specially developed software capable of recording and sending phone usage data and colocation information to a 
server located in the university premises, as described in \cite{Quin:2014}. Raw data collected by the software consists therefore 
in all call events between participating students, with timestamp and duration of the call, and timestamped colocation events,
where colocated devices were discovered through periodic Bluetooth scans performed by each smartphone.
Automated location data collection by each phone was turned off each night from 12:00~a.m. to 7:00~a.m. 

The resulting data is conveniently represented as $2$ temporal networks, the communication and the co-presence ones, in which
nodes represent students and events correspond to a phone call communication or to a co-presence event. Each communication event is
directed, represented by the calling node, the receiving node, the starting time and the duration of the call. Each co-presence event is
instead undirected, represented by two nodes, a starting time and a duration.

Each temporal network can be aggregated on any arbitrary time window. We have considered on the one hand communication and co-presence aggregated over the 
full study (one year), and on the other hand shorter periods of four months corresponding to the university terms: Term 1 (T1: May to August), Term 2 (T2: September to December) and Term 3 (T3: January to April).
Each aggregated communication network relates nodes, representing students, by directed links: a directed link
is drawn from student $i$ to student $j$ if $i$ placed at least one call to $j$ during the aggregation time window. Each
directed link can be weighted in two different ways: (1) the weight can be either the number of calls $n^c_{i\to j}$ from $i$ to $j$,
or (2) the total duration $d^c_{i\to j}$ of these calls. 
We also consider an undirected version of these communication networks
in which the weight of a link between $i$ and $j$ is simply the sum of the weights from $i$ to $j$ and from $j$ to $i$, 
$w^s_{ij} = w_{i\to j} + w_{j\to i}$ (with $w=n^c$ or $d^c$).

\begin{figure}[h!] 
\includegraphics[width=0.9\textwidth]{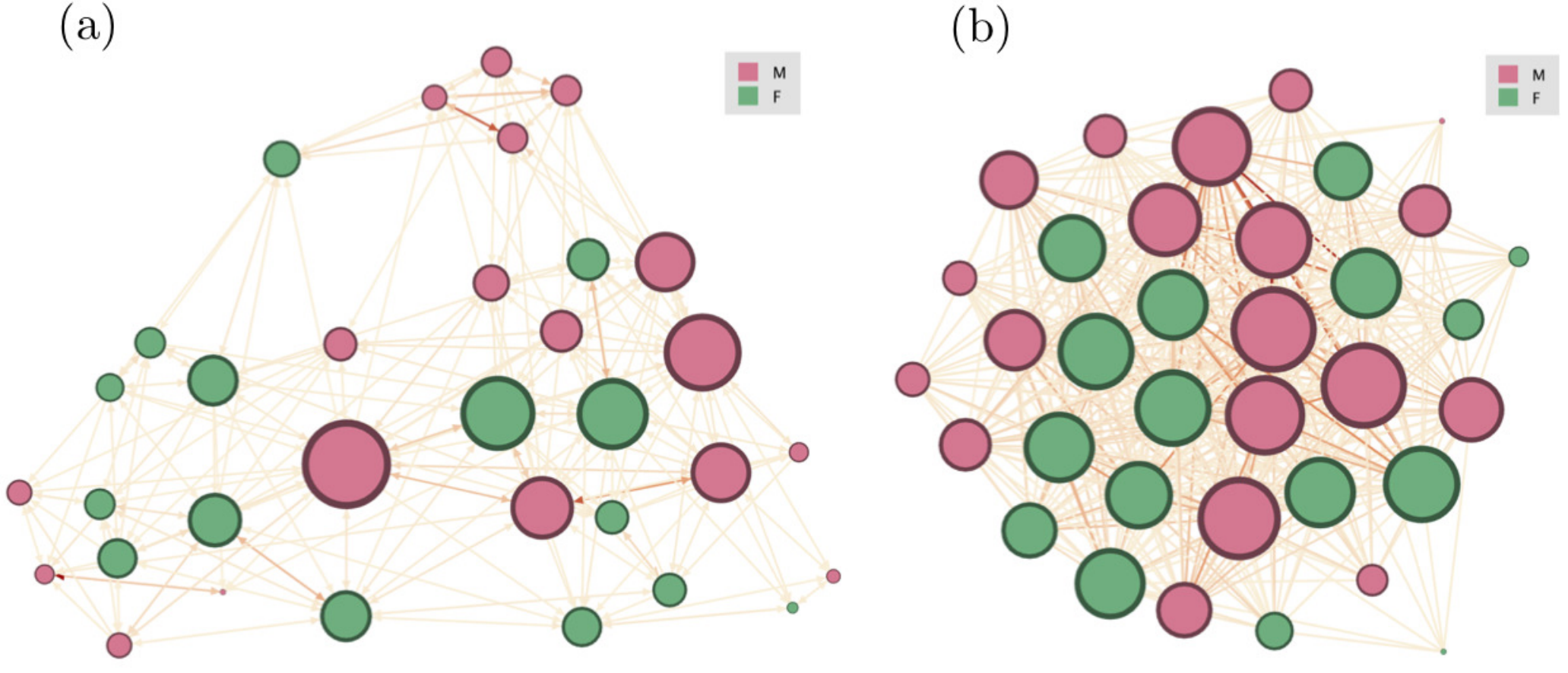}
\caption{{\bf Graphical representation of yearly aggregated networks.} 
(a) Communication network (b) Co-presence network. Nodes are colored by gender (``M'' for male and ``F'' for female), and their size depends on their degree. Edges are colored by aggregated duration.}

\label{fig_gephi}
\end{figure}

As already mentioned, the co-presence networks are undirected. Moreover, in order to discard classroom activities that are imposed by the university schedule
and not driven by personal relationships, we consider in the co-presence aggregated networks only co-presence 
events taking place either after 9:00 p.m. each day for the week days or during weekends. For each pair of students $(i,j)$, a link is drawn
if they have been detected at least once in co-presence, and the corresponding weights are defined, as in the communication network, either as the number $n^{cp}_{i j}$ of such events, or by their total duration $d^{cp}_{i j}$. Table~\ref{tab:network_properties} shows the properties of both networks under study for these time windows. Figure~\ref{fig_gephi} displays the yearly aggregated communication and co-presence networks.

In addition, questionnaires were used to assess self-reported relations among students. 
Each participant indicated his/her friendship tie-strength with all other participants by answering individually the following two questions: (Q1) ``How strong is your relationship with this person?'' and (Q2) ``How would you feel asking this friend to loan you $\$100$ or more?''. For each question, a $9$-point scale was used where $1$ indicates for Q1 that they barely know each other (resp., for Q2 that they would never ask), while $9$ indicates they are close to each other (resp., for Q2, that they would feel comfortable).
These questionnaires were answered by the students at the start of the study (T0) to establish baseline values, and subsequently at the end of every term (T1, T2, T3).
At each such time, we obtain therefore two questionnaire networks (one for each question asked). Both networks are fully-connected,
directed, and weighted, where the weight $W_{i \to j}$ of an edge from student $i$ to student $j$ ranges 
from $0.1$ to $1.0$ (9 points) indicating the reported strength of the friendship (Q1) or trust (Q2) relationship of $i$ towards $j$.

Finally, several attributes are available for each student: (1) his/her so-called cohort class (the students were divided into two cohort classes of approximately 50 students), (2) gender, (3) nationality, (4) first spoken language (all students can be considered bilingual to a certain extent, with some participants being fluent in three or more languages), (5) academic performance measured by the participants' grade point average (GPA) in each term. Table~\ref{tab:demography} summarizes the demographic composition of participants in terms of gender and nationality.
Self-reported data about psychological factors such as loneliness, classroom community, and adaptation to college life were also collected by means of a questionnaire at the end of each term. For each psychological factor surveyed, a numerical index was used (see~\cite{Quin:2014} for details): 
\textit{(i)} The UCLA loneliness scale (LS) ranges from a minimum of 20 to a maximum of 80, where a higher score indicates a greater sense of loneliness;
\textit{(ii)} The classroom community scale (CC) consists of 20 items that measure the individual sense of community in a learning environment,
leading to a total score ranging between 0 and 40, with a higher score indicating a greater sense of community; \textit{(iii)} The student adaptation to college questionnaire (SACQ) was applied to measure college adjustment, with higher scores indicating better adjustment.

For each attribute, the population under study was divided into two groups. For gender and cohort class, the division is straightforward. For nationality, the participants were divided into two groups---Singaporeans and foreigners---although several nationalities are represented (see Tab.~\ref{tab:demography}). With respect to the
first spoken language, in order to avoid confounding effects
with respect to nationality, we focus only on Singaporean students, whose first language is either English or Chinese.
For academic performance (GPA) and the psychological indices, again the participants were segregated into two groups to facilitate the analysis of the results: first group with above-the-median values, and the other group with below-the-median values.
\begin{table}[h!]
\caption{\bf Properties of the communication and co-presence networks.}
\small
\centering
\begin{subtable}{1\textwidth}
\caption{\bf Aggregation period: 1 year}
 \centering
\begin{tabular}{p{4.5cm}|p{2.5cm}|p{2.5cm}}
\hline
\textbf{Network Properties} & \textbf{Communication} & \textbf{Co-presence}\\
\hline
\hline
Number of nodes & 33& 35 \\
Number of edges& 282 &435 \\
Average degree &8.5 &24.85\\
Average clustering coefficient& 0.45& 0.858\\
Edge weight -- Total number& 2,125 &14,249\\
Edge weight -- Average number&64.4 &407.1\\
Edge weight -- Total duration (s)& 93,580 &11,966,878\\
Edge weight -- Average duration (s) &2,835.8& 341,911.0\\
\hline
\end{tabular}
\label{table:1}
\end{subtable}
\begin{subtable}{1\textwidth}
\caption{\bf Aggregation period: 4 months (Term)}
\centering 
\begin{tabular}{p{4.4cm}|p{0.8cm}|p{0.8cm}|p{0.8cm}|p{1.2cm}|p{1.2cm}|p{1.2cm}}
\hline
\multirow{2}{*}{\bf Network Properties} & \multicolumn{3}{c|} {\bf Communication} & \multicolumn{3}{c}{\bf Co-presence} \\
\cline{2-7}
& \centering T1 & \centering T2 & \centering T3 & \centering T1 & \centering T2 & {\centering{T3}} \\
\hline
\hline
Number of nodes&  33 & 33 & 33&  35&  34&  31 \\
Number of edges&  162&  150&  129&  347 &292 & 145 \\
Average degree&  9.81 & 9.09&  7.81 & 19.82&  17.17 & 9.35 \\
Average Clustering coefficient&  0.30 & 0.34 & 0.23&  0.80 & 0.80&  0.65 \\
Edge weight -- Total number&  766 & 822&  555 & 5387&  7057&  1791 \\
Edge weight -- Average number&  23.2 & 24.9 & 16.8&  153.9&  207.5&  57.8 \\
Edge weight -- Total duration (s)&  29,969&  37,553&  26,550&  3,981,721&  6,412,726 & 1,493,623 \\ 
Edge weight -- Average duration (s)&  908.1&  1,138&  804.5&  11,3763.4&  188,609.5& 48,181.3\\
\hline
\end{tabular}
\label{table:2}
\end{subtable}
\label{tab:network_properties}
\end{table}


\begin{table}[h!]
\small
\centering 
\caption{{\bf Demography Table:} Number of participants by nationality and gender.}
\begin{tabular}{p{5.5cm}|p{1cm}|p{1cm}|p{1cm}}
\hline
\multirow{2}{*}{\bf Nationality} & \multicolumn{2}{c|} {\bf Gender} & \multirow{2}{*}{\bf Total} \\
\cline{2-3}
& \textit{Male} & \textit{Female} \\
\hline
\hline
Singaporean/Permanent Resident (PR) & 7 &15& 22 \\
Indian& 2& 0 &2\\
PRC Chinese &4& 1& 5\\
Malaysian &4& 0& 4\\
Vietnamese& 2& 0 &2\\ 
{\bf Total} &\textbf{19} &\textbf{16} &\textbf{35}\\
\hline
\end{tabular}
\label{tab:demography}
\end{table}

\subsection*{Measuring homophily}

Homophily in a social network can be assessed in a number of ways. It is possible for instance to investigate the fraction of ties between individuals with similar
versus different characteristics, but also higher-order structures such as triads~\cite{Laniado:2016}, and even temporal patterns or motifs~\cite{Kovanen:2013}. Given the weighted nature of the networks at hand---with possibly broad distributions
of weights as often encountered in human interaction networks, taking into consideration edge weights is crucial~\cite{Mollgaard:2016}.

Here, we consider the following metrics to describe and quantify homophily in each network, and for each node attribute $A$:
\begin{itemize}
\item \textbf{Dyadic homophily:} we first consider homophily at the basic dyad level, i.e., considering the basic elements forming the network, that is the edge. We compute the total fraction of weights carried by edges between nodes with the same value of the attribute $A$ (directed networks being converted to their undirected versions):
\begin{equation}
D = \frac{\sum_{i,j / A_i = A_j} w_{ij}}{\sum_{i,j} w_{ij}}  .
\end{equation}
 
\item \textbf{Triadic homophily:} closed triangles describe the smallest non-trivial structure in a social network. For a given attribute $A$, that 
can take only two values, triangles can either be formed by three individuals with equal value of the attribute, or by a group of $2$ individuals
different from the third. We therefore compute the ratio of the weights of triangles formed by individuals with the same attribute value to the
total weight carried by triangles:
\begin{equation}
T = \frac{\sum^\Delta_{i,j,k / A_i = A_j = A_k}  ( w_{ij} + w_{ik} + w_{jk} ) }{\sum^\Delta_{i,j,k} (w_{ij} + w_{ik} + w_{jk})}  ,
\end{equation}
where the sums $\sum^\Delta$ are conditioned on $ijk$ being a closed triangle.
To compute this index, we convert directed networks to their undirected versions.

\item \textbf{Social preference:} for each node $i$, we can rank his/her neighbors $j$ according to 
the value of the corresponding edge weight $w_{i \to j}$. We then compare the attributes of $i$ and of his/her first-ranked neighbor and compute the fraction of individuals for which these attributes are equal. We compute moreover these fractions
separately for all nodes $i$ with a given value of the attribute $A$. For instance, we can compute separately the fraction of male students
and of female students for whom the strongest link is towards a male student.
 
\item \textbf{Temporal motifs:} as put forward in~\cite{Kovanen:2013}, the availability of time-resolved data makes it possible to investigate
homophily in temporal patterns of interactions by considering events concerning the same set of nodes and close enough in time. As in~\cite{Kovanen:2013}, we consider sets of events separated by at most $10$ minutes and involving the same $2$ or $3$ individuals, and
 investigate the similarity (or difference) of their attributes. For the sake of simplicity and given the lack of statistics for motifs involving more than 
 $2$ nodes in our data, we limit the evidence shown to reciprocal and repeated calls (within the time-window of $10$ minutes) between
 two nodes: we consider all such patterns and compute the fraction involving nodes with equal attributes.
\end{itemize}

\noindent
\textbf{Null model:} 
The existence of homophily in each of the above-defined measures needs to be assessed by means of comparison with a proper null model. Here, we consider as null model
a random reshuffling of the attributes among the nodes. This procedure indeed keeps the network structure intact. We perform $100$ reshuffling
and compute the homophily indices for each. The empirical value is  compared to the resulting distribution (shown in figures as a boxplot, with the box extremities representing the $25^{\mathrm{th}}$ and $75^{\mathrm{th}}$ percentiles of the distributions, and whiskers
at the $5^{\mathrm{th}}$, $10^{\mathrm{th}}$, $90^{\mathrm{th}}$ and $95^{\mathrm{th}}$ percentiles). It is considered that the data reveals an absence of homophily if the data point falls within the box (``No''), and that we have respectively weak (``W''), strong (``S'') and very strong (``VS") degrees of homophily if the data point lies 
respectively between the $75^{\mathrm{th}}$ and the  $90^{\mathrm{th}}$ percentiles, between the $90^{\mathrm{th}}$ and the $95^{\mathrm{th}}$ percentiles, and above the $95^{\mathrm{th}}$ percentile. In addition, we find in few cases evidence for heterophily, 
i.e., the tendency to have less homophilic dyads, triads or motifs with respect to the null model. Similarly to the homophily patterns, we 
consider that we have respectively weak  (``W$_\mathrm{het}$''), 
strong (``S$_\mathrm{het}$'') and very strong (``VS$_\mathrm{het}$'') degrees of heterophily when the data point lies 
respectively between the $10^{\mathrm{th}}$ and the  $25^{\mathrm{th}}$ percentiles, between the $5^{\mathrm{th}}$ and the $10^{\mathrm{th}}$ percentiles, and below the $5^{\mathrm{th}}$ percentile of the null model distribution.

Finally, and for the sake of simplicity, we will also envision a coarser classification of patterns, in which we group the cases 
``W'', ``No'' and ``W$_\mathrm{het}$'' together (and as no evidence for 
homophily nor heterophily), and we consider as evidence for homophily
(resp. heterophily)
both ``S'' and ``VS'' cases (resp. ``S$_\mathrm{het}$'' and ``VS$_\mathrm{het}$'').

\subsection*{Networks comparison}

The data at hand defines different types of relationship among students: specifically, communication, co-presence, friendship and trust relations. It is worth noting that these data are available with different temporal resolutions throughout the 12-month study period. To enable a meaningful comparison of these networks, we resort to two distinct metrics:
\begin{itemize}
\item The {\em Pearson correlation coefficient} between the weights of links between individuals within the two considered
networks. If one of the network is directed and the other undirected, we first convert the directed one into its undirected counterpart: for each pair of nodes $(i,j)$, the resulting weight is the sum of the weights on the directed edges $i\to j$ and $j\to i$.
The correlation values
are described as: very weak if $\in [0; 0.19]$, weak if $\in [0.2; 0.39]$, moderate if $\in [0.4; 0.59]$,
strong if $\in [0.6; 0.79]$, very strong if $\in [0.8 ; 1]$ (see Ref.~\cite{evans1996straightforward}).
\item The {\em cosine similarity} for each node $i$, which measures the similarity between this node and its neighborhoods in the two networks. If $w_{ij,1}$
and $w_{ij,2}$ denote the weights on the links from $i$ to $j$ respectively in networks $1$ and $2$, the cosine similarity of $i$ is defined
as 
\begin{equation}
\text{sim}_{1,2}(i) = \frac{\sum_j w_{ij,1} w_{ij,2} } { \sqrt{\sum_j w_{ij,1}^2}   \sqrt{\sum_j w_{ij,2}^2 } }  .
\end{equation}
We compute the distribution of $\text{sim}_{1,2}(i)$ for a pair of networks and compare it with two null models: in the first one, 
we keep the link structure and reshuffle the weights on the links; in the second, we reshuffle the links while keeping the degree of each node fixed~\cite{Maslov:2004}.
\end{itemize}

While these measures give us an idea of the topological similarity of networks, our goal here is also to provide a way to estimate whether homophily patterns are exhibited consistently across different networks.
To this aim, we tabulate for each network and each homophily index used---e.g., dyadic homophily, triadic homophily, etc.---the number of occurrences corresponding to an absence of homophily, weak, strong, or very strong evidence of homophily
(or heterophily). 
We then compute the number
of concordant and discordant cases for each pair of networks. For instance, we track the number of indices for which no evidence of homophily is found in one network, while strong evidence is uncovered in the second network. This gives us a first indication with respect to whether homophily patterns are similar across two networks. Moreover, we compare these numbers to a null model defined as follows: for each network
and each homophily index, we reshuffle the 
``No'', ``W'', ``S'', ``VS'',
``W$_\mathrm{het}$'', ``S$_\mathrm{het}$'', ``VS$_\mathrm{het}$''
cases, keeping their number fixed, and compute 
again the number of concordant and discordant indices. If the empirical number of concordant cases falls outside the confidence interval
of the resulting distribution for the null model, it indicates that the number of concordant cases obtained is not just due for instance to a large
majority of ``No" cases. Thus, it is a strong indication that the homophily patterns between networks are similar enough so that information on homophily can be obtained from either.

\section*{Results}

\subsection*{Description of network characteristics}

We first present an overview of some descriptive characteristics of the data under investigation. 

\begin{figure}[h!]
\includegraphics[width=0.9\textwidth]{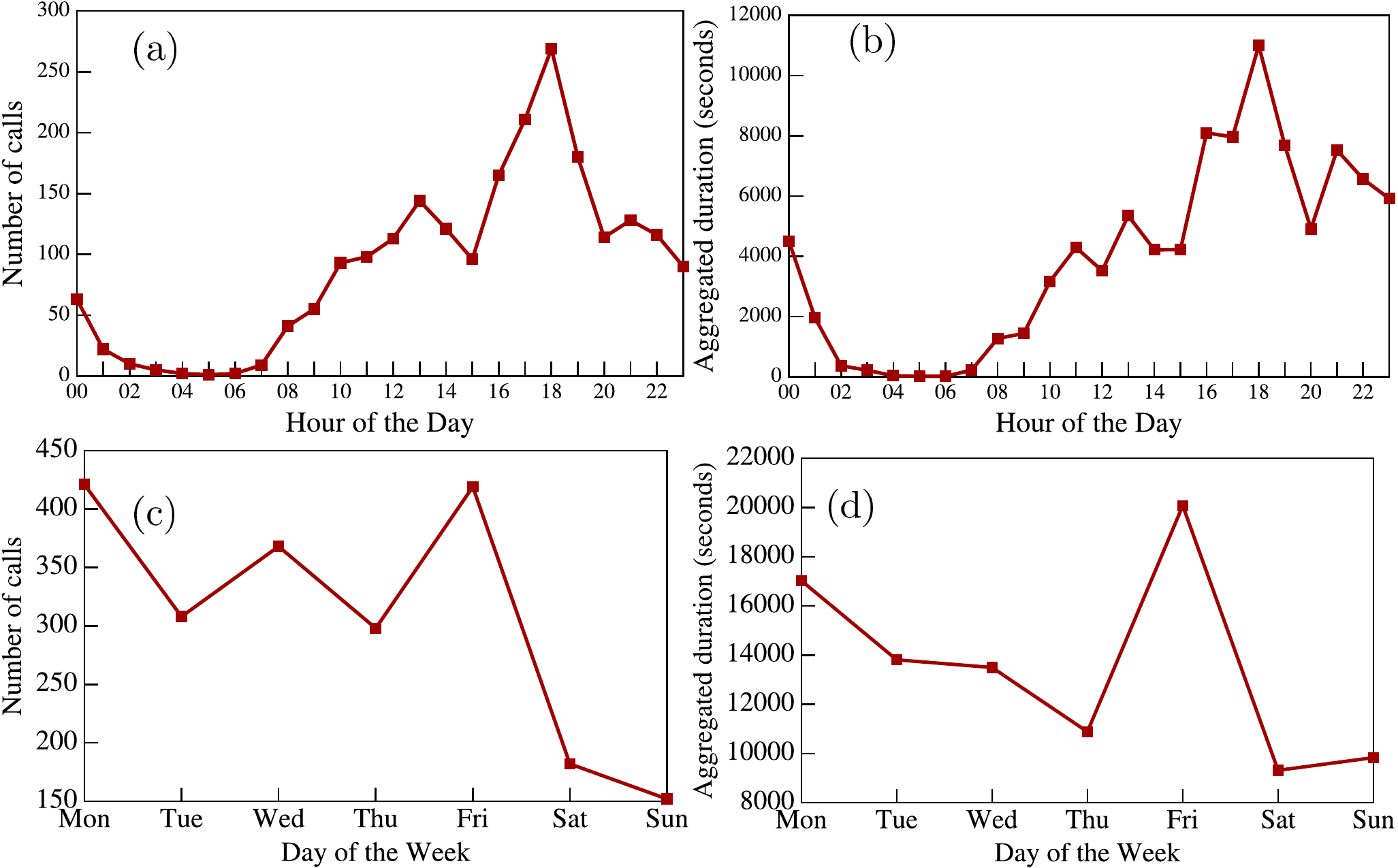}
\caption{{\bf Distribution of communication events over time.} (a) Number of calls between participants on an hourly basis throughout the day; (b) Aggregated duration of calls on an hourly basis throughout the day; (c) Number of calls on a daily basis throughout the week; (d) Aggregated duration of calls on a daily basis throughout the week.}

\label{figtmcall}
\end{figure}

\begin{figure}[h!]
  \includegraphics[width=0.9\textwidth]{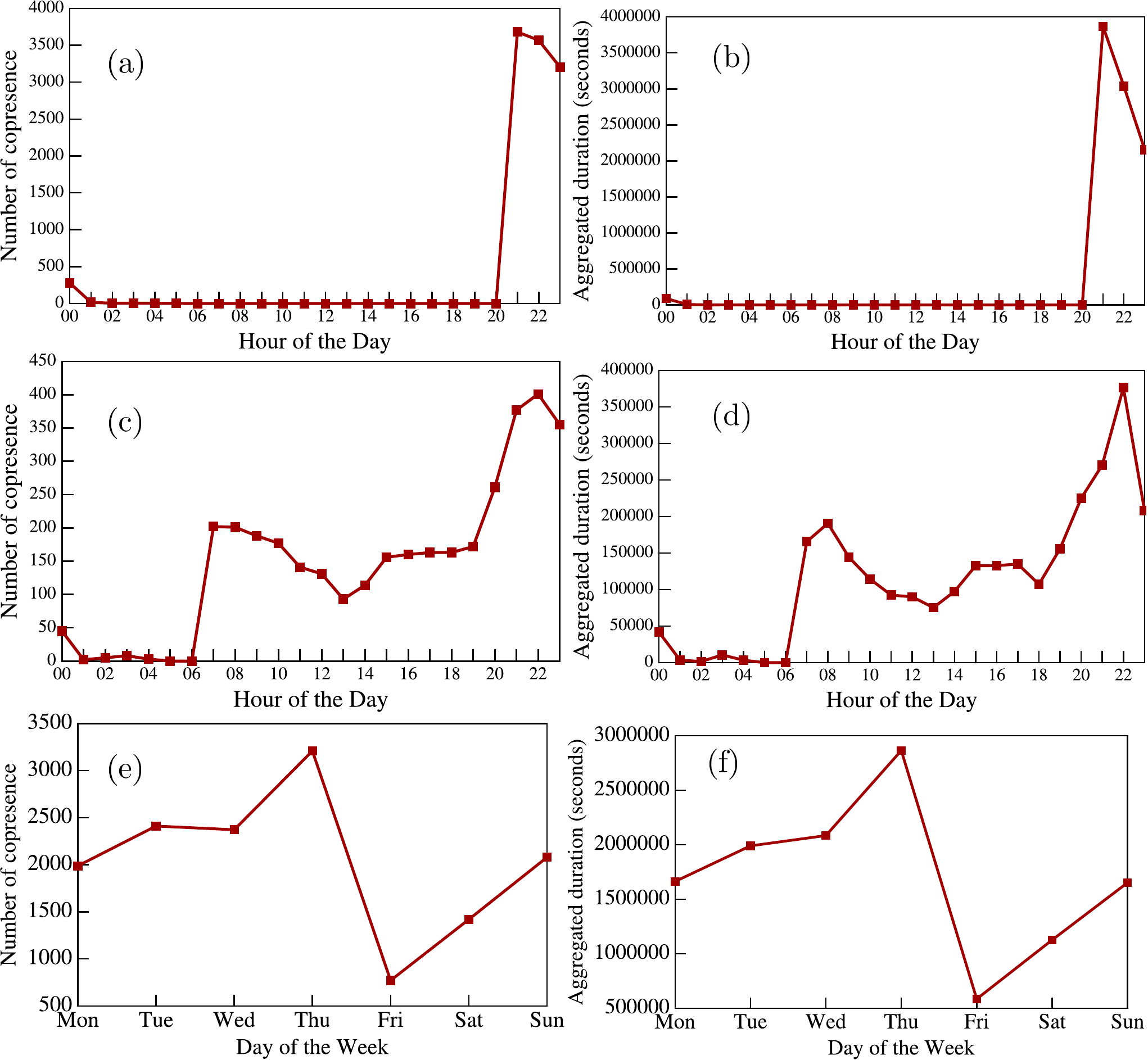}
\caption{{\bf Distribution of co-presence events over time.} (a) Number of co-presence events throughout the day on weekdays; (b) Aggregated duration of co-presence events throughout the day on weekdays; (c) Number of co-presence events throughout the day on weekends; (d) Aggregated duration of co-presence events throughout the day on weekends; (e) Number of co-presence events on a daily basis throughout the week; (f) Aggregated duration of co-presence events on a daily basis throughout the week. 
Co-presence network: Monday to Friday -- 9~p.m. to 12~a.m., Saturday, Sunday -- 6~a.m. to 12~a.m.
}
\label{figtmcopr}
\end{figure}

Figures~\ref{figtmcall} and~\ref{figtmcopr} show the distribution of calls and co-presence events as a function of the hour of the day, summed over all days of data collection, and as a function of the day of the week, 
summed over all weeks. As expected, communication events display clear daily and weekly patterns, with 
almost no calls at night, an increase during the day, and a peak around $6-7$~p.m. around the end of class time. It is worth adding that all participants dwelled on campus from Monday to Friday as part of their residential program requirements. Few calls were placed during weekends, with instead more calls on Fridays and Mondays. Interestingly, co-presence events exhibit instead a peak on Thursdays, which may be attributed to the fact most Singaporean students leaved the campus on Friday evenings. During weekends, co-presence peaks in the evenings, especially on Sunday when students come back to stay on campus in preparation for school the next day. 

\begin{figure}[h!]
  \includegraphics[width=0.9\textwidth]{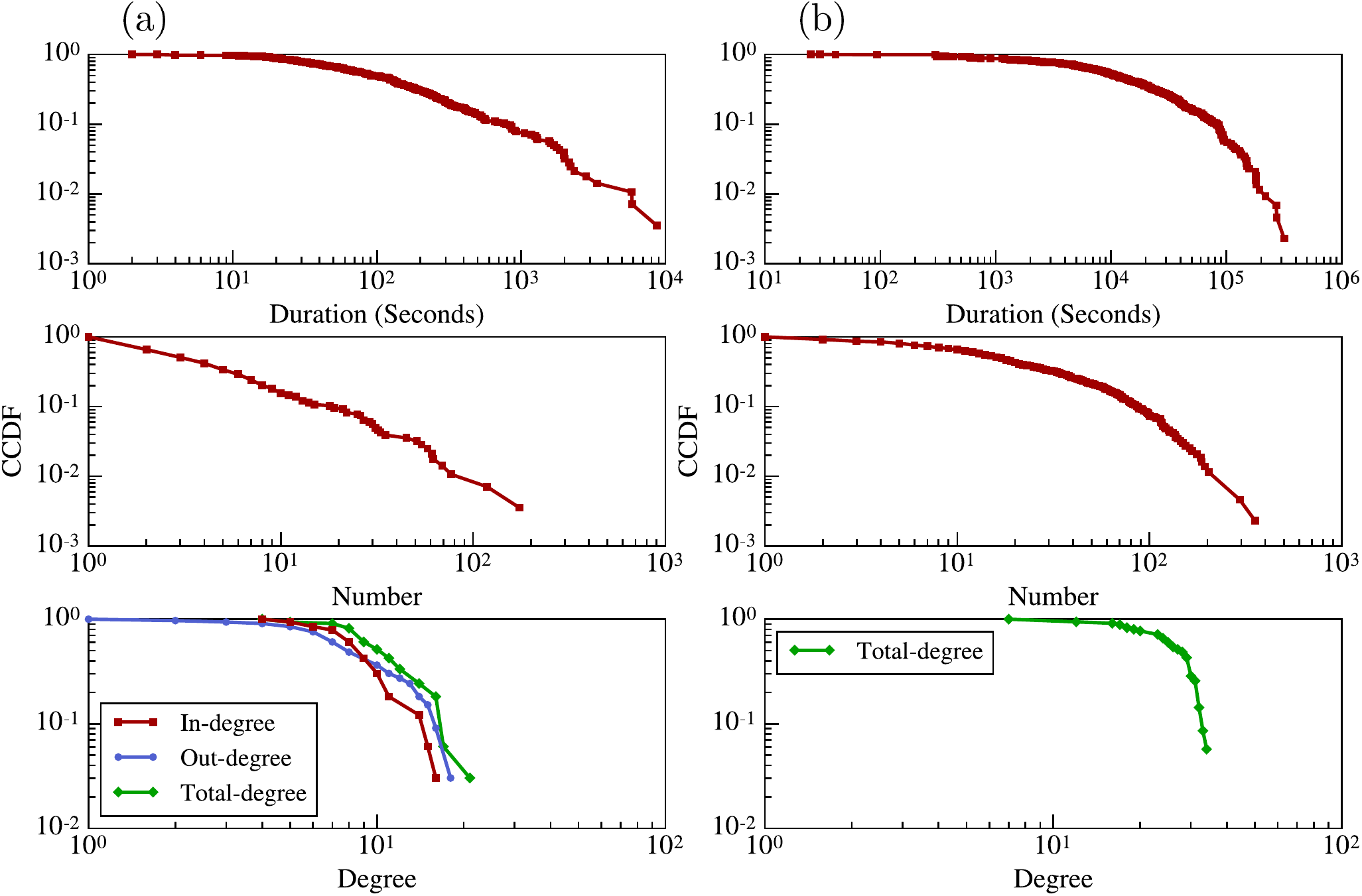}
  \caption{{\bf Complementary cumulative distributions of edge weights.} top: weights given by the number of events; middle: weights given by the aggregated durations; bottom: node degrees for yearly-aggregated networks. (a) Communication network; (b) Co-presence network. }

\label{figccdf}
\end{figure}

\begin{figure}[htbp]
  \includegraphics[width=0.9\textwidth]{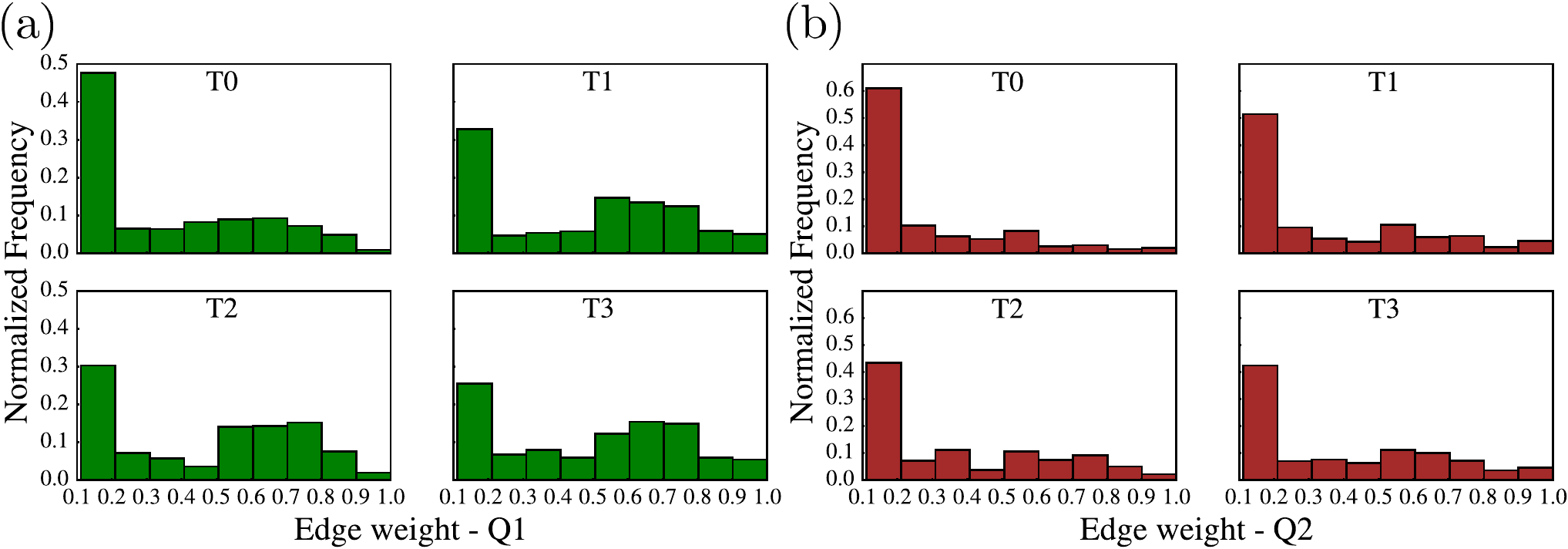}
\caption{{\bf Histograms of weights for the networks defined by the two questionnaires.} (a) Q1; (b) Q2.}

\label{figpdfQ}
\end{figure}

Figure~\ref{figccdf} shows the complementary cumulative distribution (CCDF) of edge weights (number and aggregated durations of events) and of node degrees
(number of neighbors of a node) in the yearly aggregated communication and co-presence networks. As expected in this type of networks,
edge weights show broad distributions spanning several orders of magnitude. On the other hand, node degree distributions are narrow as the population
under investigation is of relatively small size ($35$ students). We note that, even considering yearly aggregation, the networks are far from being
fully connected, especially for the communication network: each student had on average communicated only with less than $10$ other students, and the maximal degree is $22$, in line with
results on limited communication capacities observed
in larger systems \cite{Miritello:2013}.
Finally, Fig.~\ref{figpdfQ} displays the distribution of weights in the questionnaire networks. Most links carry the minimum possible weight in all cases, but this tendency decreases over time in both questions (see Data and Methods for the exact phrasing), while the fraction of strong friendships tends to increase, and the distribution tends towards a bimodal shape.
\subsubsection*{Comparison between successive terms}

\begin{figure}[h!]
  \includegraphics[width=0.9\textwidth]{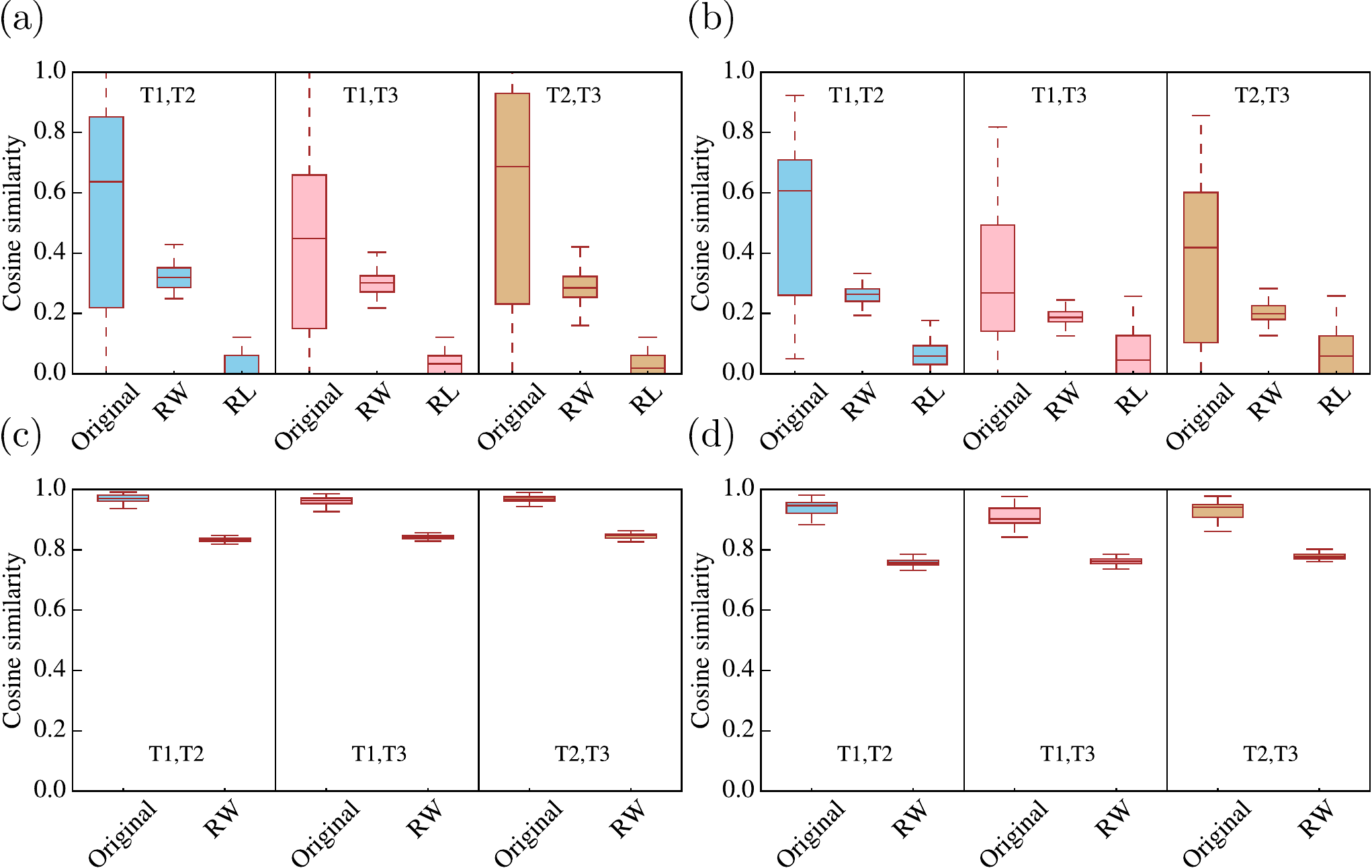}
\caption{{\bf Boxplots of the cosine similarity distributions within the networks aggregated in different terms (T1, T2, T3),
compared with the same distributions for networks with reshuffled weights (RW) or links (RL).} (a) Communication networks -- weights given by aggregated call durations; (b) Co-presence networks -- weights given by aggregated durations; (c) Questionnaire -- Q1; (d) Questionnaire -- Q2. For the questionnaire cases 
the RL and RW procedures are equivalent as the
networks are fully connected.
}

\label{figcs}
\end{figure}

Table~\ref{table:pearsons} and Fig.~\ref{figcs} illustrate the temporal evolution of the different networks at the term level. 
The communication networks aggregated in the second and third terms are very strongly correlated, while they are only
moderately correlated with the first term network. On the other hand,
the co-presence networks in different terms show weak to moderate correlations.
For both networks, the cosine similarity distribution extends
over a quite broad range (Fig.~\ref{figcs}), and 
show larger values than in the two null models considered, with lowest median value for the similarities
between the non-successive terms T1 and T3.  
Finally, for each type of questionnaire question, the correlation between
the weights decrease as the time between questionnaires increases. In particular, the network constructed from the questionnaire answered
at the start of the study shows the weakest correlation with successive questionnaires, which may be attributed to the fact that the students did
not know each other well at that stage. Cosine similarities
between different terms take very large values, much larger than in the null model with reshuffled weights (Fig.~\ref{figcs}).
\begin{table}[h!]
\caption{\bf Pearson correlation coefficients between term-aggregated networks}
\small
\centering
\begin{subtable}{1\textwidth}
\caption{{\bf Communication vs. communication.} ($p=0$ for all entries)} 
\centering 
\begin{tabular}{ |p{1.8cm}|p{1.8cm}|p{1.8cm}|p{1.8cm}||p{1.7cm}|p{1.7cm}|p{1.7cm}|   }
\hline
&Duration-T1& Duration-T2 & Duration-T3 & 
Number-T1&Number-T2 & Number-T3 \\
\hline
Duration-T1 & 1.0 & 0.57 & 0.59 & 0.89 &0.49 & 0.59 \\
Duration-T2& 0.57 & 1.0 & 0.89 &0.6 &0.9 &0.88\\
Duration-T3 &0.59 &0.89 &1.0 & 0.52 & 0.77 &0.9 \\
\hline \hline
Number-T1 &0.89 & 0.6 &0.52 &1.0 & 0.56 &0.57 \\
Number-T2 &0.49 & 0.9 & 0.77 &0.56 & 1.0  &0.9\\
Number-T3 &0.59 &0.88 &0.9 & 0.57 & 0.9 &1.0 \\
\hline
\end{tabular}
\label{table:4}
\end{subtable}
\\[3ex]
\begin{subtable}{1\textwidth}
\caption{\bf Co-presence vs. co-presence}
\centering 
\begin{tabular}{ |p{1.8cm}|p{1.9cm}|p{1.9cm}|p{1.9cm}||p{1.9cm}|p{1.9cm}|p{1.9cm}|   }
\hline
&Duration-T1& Duration-T2 & Duration-T3 & 
Number-T1&Number-T2 & Number-T3 \\
\hline
Duration-T1 &1.0 $(p=0.0)$& 0.4 $(p=0.0)$& 0.44 $(p=0.0)$& 0.93 $(p=0.0)$ &0.37 $(p=0.0)$& 0.2 (p=0.04)\\
Duration-T2 &0.4 $(p=0.0)$& 1.0 $(p=0.0)$& 0.41 $(p=0.0)$& 0.32 $(p=0.0)$& 0.91 $(p=0.0)$ &0.3 (p=0.002)\\
Duration-T3& 0.44 $(p=0.0)$& 0.41 $(p=0.0)$& 1.0 $(p=0.0)$& 0.48 $(p=0.0)$& 0.43 $(p=0.0)$& 0.87 $(p=0.0)$\\
\hline
\hline
Number-T1& 0.93 $(p=0.0)$& 0.32 $(p=0.0)$& 0.48 $(p=0.0)$& 1.0 $(p=0.0)$& 0.35 $(p=0.0)$& 0.3 (p=0.002)\\
Number-T2& 0.37 $(p=0.0)$& 0.91 $(p=0.0)$& 0.43 $(p=0.0)$& 0.35 $(p=0.0)$& 1.0 $(p=0.0)$& 0.44 $(p=0.0)$\\
Number-T3& 0.2 (p=0.04)& 0.3 (p=0.002)& 0.87 $(p=0.0)$& 0.3 (p=0.002)& 0.44 $(p=0.0)$& 1.0 $(p=0.0)$\\
\hline
\end{tabular}
\label{table:5}
\end{subtable}
\\[3ex]
\begin{subtable}{1\textwidth}
\caption{{\bf Questionnaire vs. questionnaire.} ($p=0$ for all entries)}
\centering 
\begin{tabular}{ |p{1cm}|p{1.cm}|p{1.cm}|p{1.cm}|p{1.cm}||p{1.cm}|p{1.cm}| p{1.cm}|p{1.cm}|  }
\hline
&Q1-T0&Q1-T1& Q1-T2 & Q1-T3 & 
Q2-T0&Q2-T1&Q2-T2 & Q2-T3 \\
\hline
Q1-T0 & 1.0 & 0.65 & 0.58 & 0.51 & 0.62 & 0.55 & 0.52 & 0.43 \\
Q1-T1 & 0.65 & 1.0 & 0.81 & 0.77 & 0.47 & 0.71 & 0.7 & 0.63 \\
Q1-T2 & 0.58 & 0.81 & 1.0 & 0.77 & 0.37 & 0.6 & 0.73 & 0.59\\
Q1-T3& 0.51 & 0.77 & 0.77 & 1.0 & 0.28 & 0.52 & 0.63 & 0.73 \\
\hline \hline
Q2-T0& 0.62 & 0.47 & 0.37 & 0.28 & 1.0 & 0.58 & 0.49 & 0.34 \\
Q2-T1& 0.55 & 0.71 & 0.6 & 0.52 & 0.58 & 1.0 & 0.71 & 0.59 \\
Q2-T2& 0.52 & 0.7 & 0.73 & 0.63 & 0.49 & 0.71 & 1.0 & 0.65\\
Q2-T3& 0.43 & 0.63 & 0.59 & 0.73 & 0.34 & 0.59 & 0.65 & 1.0\\
\hline
\end{tabular}
\label{table:6}
\end{subtable}
\label{table:pearsons}
\end{table}

\subsubsection*{Comparison between the communication, co-presence, friendship and trust networks}

We found no significant correlation between the weights of edges in the yearly- or term-aggregated communication and co-presence networks,
showing that these networks correspond potentially to quite different interaction patterns 
(the cosine similarities between these networks show also
quite low values). On the other hand, both communication and co-presence weights show weak but significant correlations with 
the weights resulting from the two questionnaires Q1 and Q2. The values of the cosine similarities of neighborhoods of nodes (i) between communication and questionnaires, and 
(ii) between co-presence and questionnaires, display moreover
values much larger than in the null models with reshuffled weights or edges. Finally, in each term, 
the weights reported in Q1 and Q2 are strongly correlated (but distinct), and the cosine similarities
of neighborhoods of nodes in the two questionnaire networks are close to $1$ (see Supporting Information). 

\begin{figure}[h!]
  \includegraphics[width=0.9\textwidth]{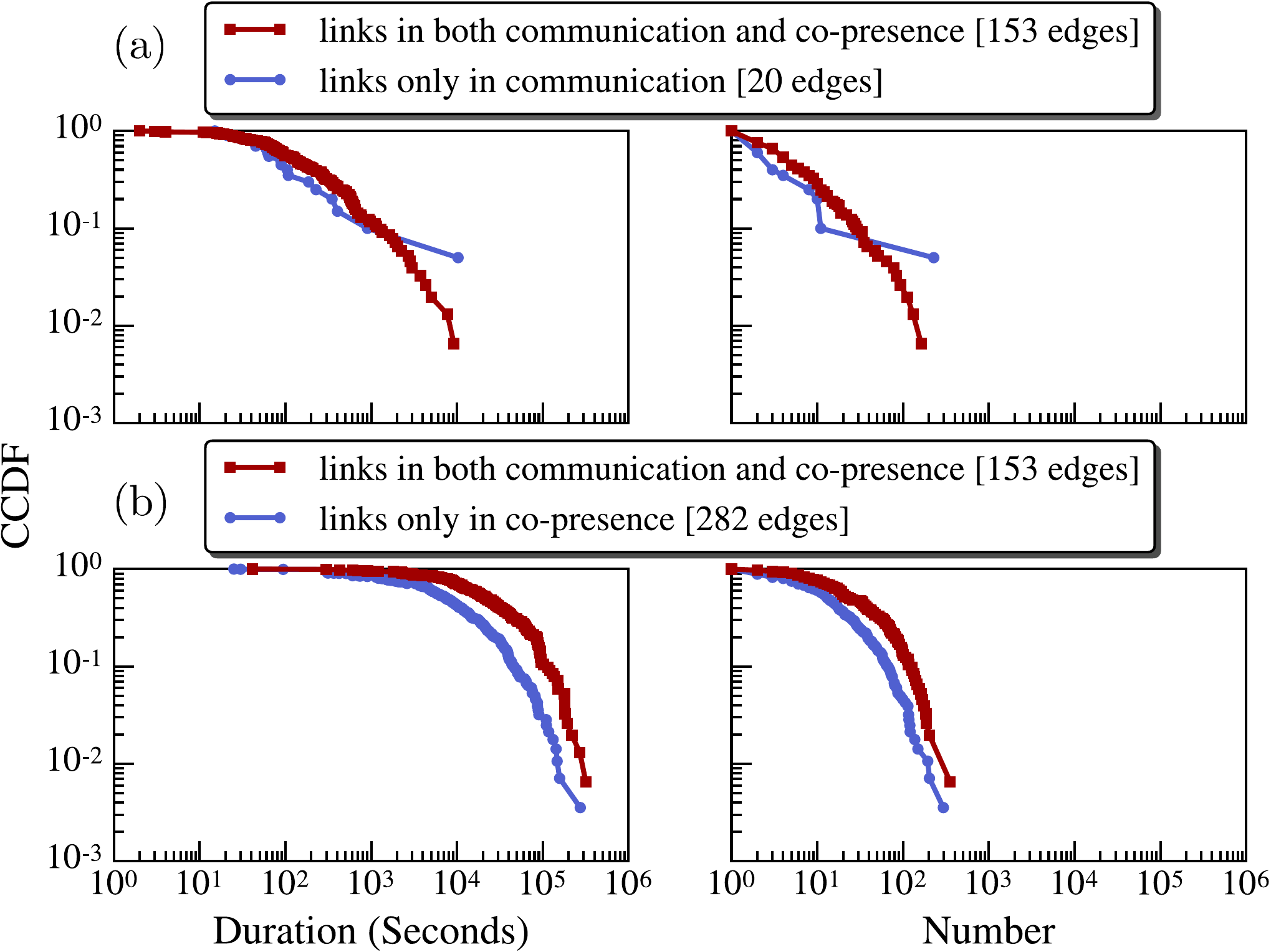}
\caption{{\bf CCDF of weights of links common to communication and co-presence networks, compared
to the CCDF of weights of links present only in one of these networks.} (a) CCDF of the weights in the communication network; (b) CCDF of the weights in the co-presence network.
} 

\label{fig11}
\end{figure}

To explore in more details the comparison between pairs of networks,  
we consider the properties of links either {\em (i)} common to two networks or {\em (ii)} present only in one of two networks. Figure~\ref{fig11} displays the CCDF of edge weights for links common to the communication and co-presence networks, as well as the CCDF 
of weights for links present in only one of the two networks. Note that many links are present only in the co-presence network, while few
are present only in the call network, which is not surprising given the much denser nature of the co-presence network.
 A clear difference is observed between the distributions of co-presence
weights, with broader distributions for links common to both networks than for links present only in the co-presence networks: students
who communicated by phone calls also tended to spend more time in co-presence, but a broad distribution is obtained even
for the links between students who did not communicate by phone. On the other hand, no clear difference is observed 
in the communication weights between pairs of students who were at least once in co-presence and pairs who were not, maybe because
of the lack of statistics for the latter: very few pairs of students indeed communicated but were never detected in co-presence.


\begin{figure}[htbp]
  \includegraphics[width=0.9\textwidth]{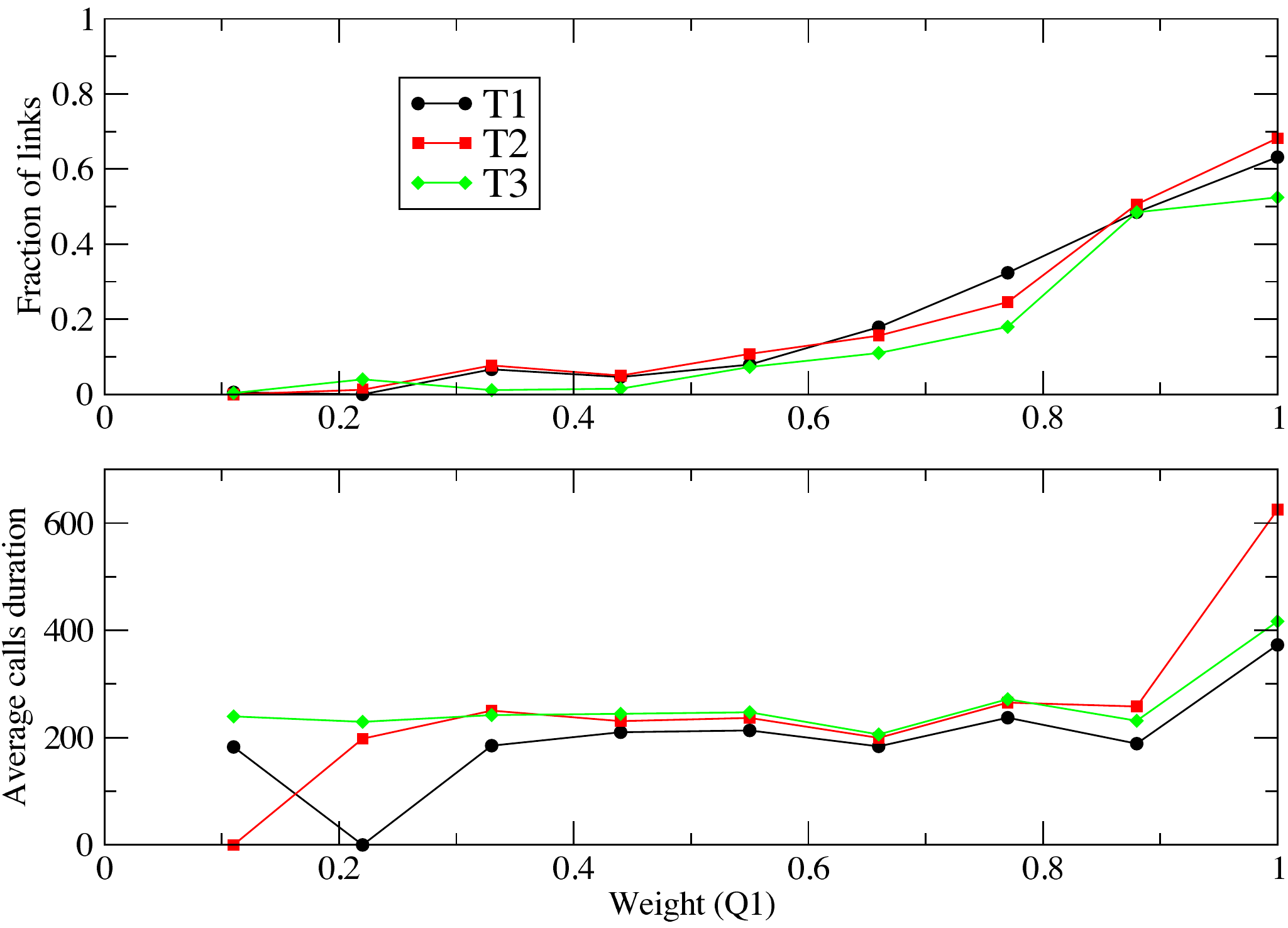}
\caption{{\bf Communication between individuals
as a function of friendship:} 
Top plot: fraction of links corresponding to a communication as a function
of their weight in Q1; bottom plot: average aggregated duration of communication along a link
(i.e., weight of the link in the communication network) as a function
of its weight in Q1.} 

\label{fig:q_w}
\end{figure}

We also compare the communication links and weights for the various weight categories in the questionnaires as shown in Fig.~\ref{fig:q_w}. 
As the questionnaire weight $w$ increases, the fraction of links with that weight that are also present in the communication network increases strongly,
from almost $0$ for low weights to $60-70\%$ for the strongest weights. This result confirms earlier findings that stronger friendship relations correspond to more probable communication.
Interestingly, however, the average number or duration of these communications does not depend on the questionnaire weight category, except for the largest
weight category, for which larger average number and duration of communications are observed: the pairs of closest friends have more frequent and longer
communication patterns with respect to other pairs of students. It is also worth highlighting that no such clear tendency is observed when comparing questionnaire weights and co-presence patterns: the fraction of links
corresponding to co-presence barely increases with the questionnaire weight, and the corresponding average co-presence duration (or number of events) does not show any clear trend (not shown).


\subsection*{Homophily patterns in yearly-aggregated networks}
 
\begin{figure}[h!]
  \includegraphics[width=0.9\textwidth]{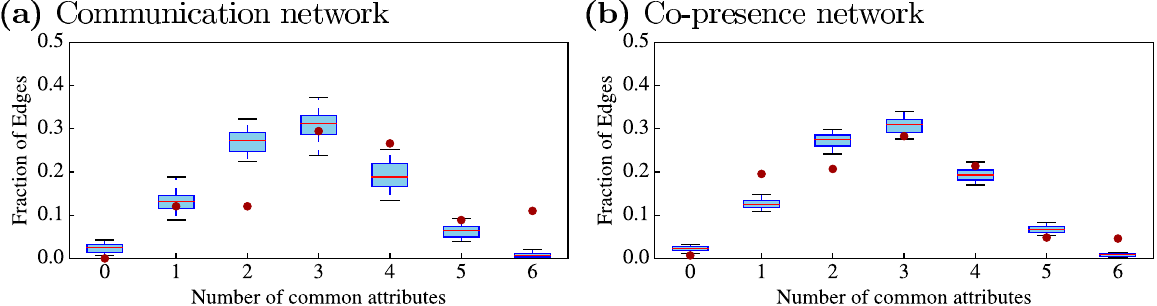}
\caption{{\bf Distribution of the number of common attributes on an edge, compared with a null model with reshuffled attributes.}
Attributes: Cohort class, Age, Gender, Nationality, GPA and First Language. 
Boxplot: Whiskers - $5^\th$, $10^\th$ and 
$90^\th$, $95^\th$ percentiles, Box - $25^\th$ and $75^\th$ percentiles.}
\label{fig1}
\end{figure}

 We first present a brief study of the homophily patterns for the globally aggregated networks. We focus here mostly on the communication network,
data for the co-presence network being shown in the 
Supporting Information. 
Figure~\ref{fig1} gives a first indication of the presence
of homophily in the communication and co-presence networks, by comparing the distribution of the number of shared attributes for individuals connected by a link
 with the same distribution in the null model in which attributes are reshuffled across nodes. Here, we consider 
 the following six attributes: cohort class, age, gender, nationality, GPA, and first spoken language. Large values of the number of shared attributes are over-represented
 with respect to the null model: in particular, a much larger fraction of links connect nodes sharing all these attributes than in the null model, while the fraction of links connecting nodes with no common attribute is smaller than in the null model.
 
\begin{figure}[h!]
  \includegraphics[width=0.9\textwidth]{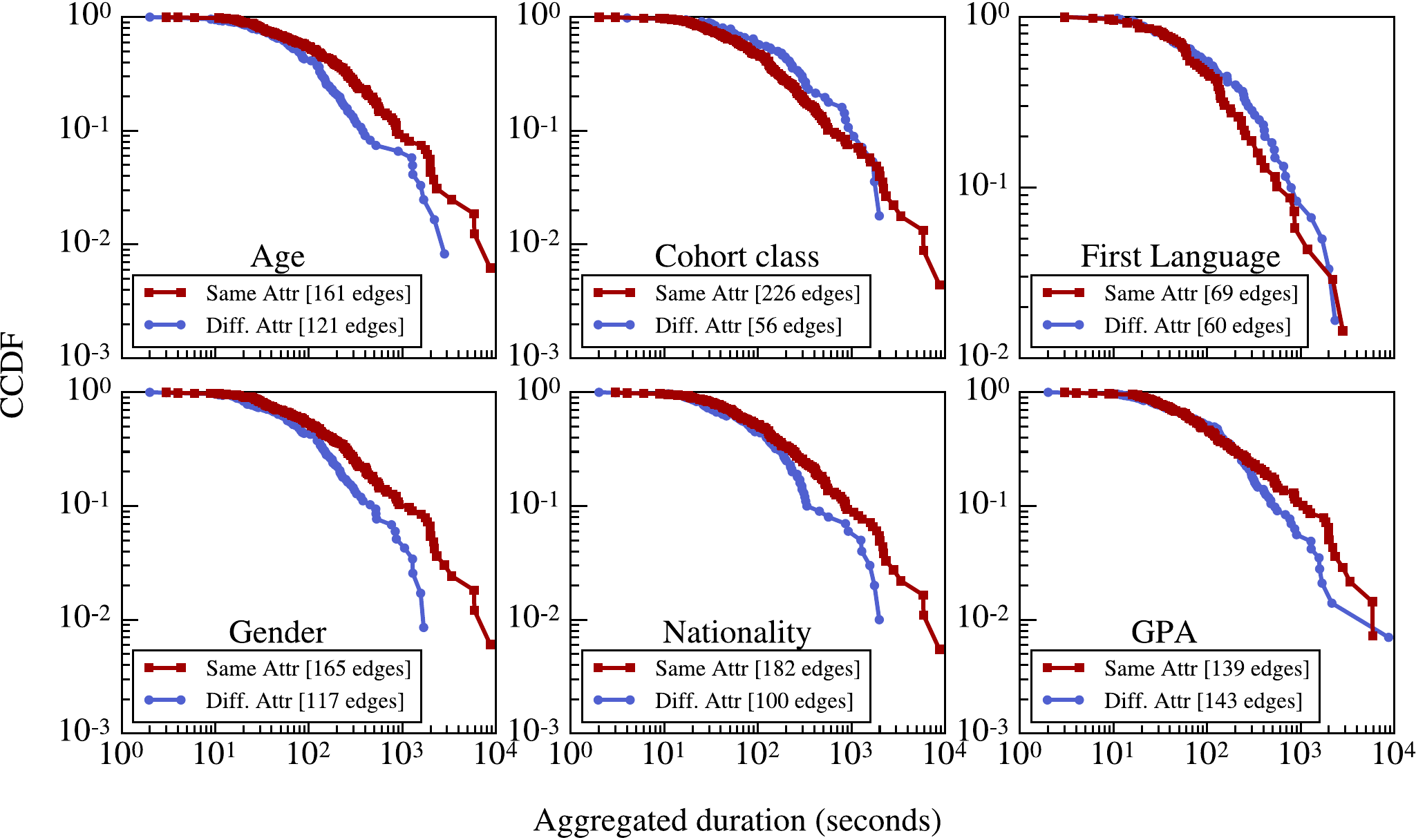}
\caption{{\bf Complementary cumulative distribution function (CCDF) of the edge weights in the yearly aggregated communication
network, for edges linking nodes
with either the same attribute or different values.} Results are shown separately for different
attributes.}

\label{fig4}
\end{figure}

Figure~\ref{fig4} goes further by showing the CCDF of edge weights in the communication network, 
separately for edges between individuals with similar and different values for
these six attributes. All distributions are broad: both weak and strong links are observed in each case, 
showing that one cannot separate these easily in two groups and guess from the weight of a link if the two connected individuals share an attribute.
On the other hand, the distributions tend to be broader for edges linking nodes with the same value for several attributes, 
and the largest weights
link nodes with same nationality, gender, age and class.

\begin{figure}[h!] 
  \includegraphics[width=0.9\textwidth]{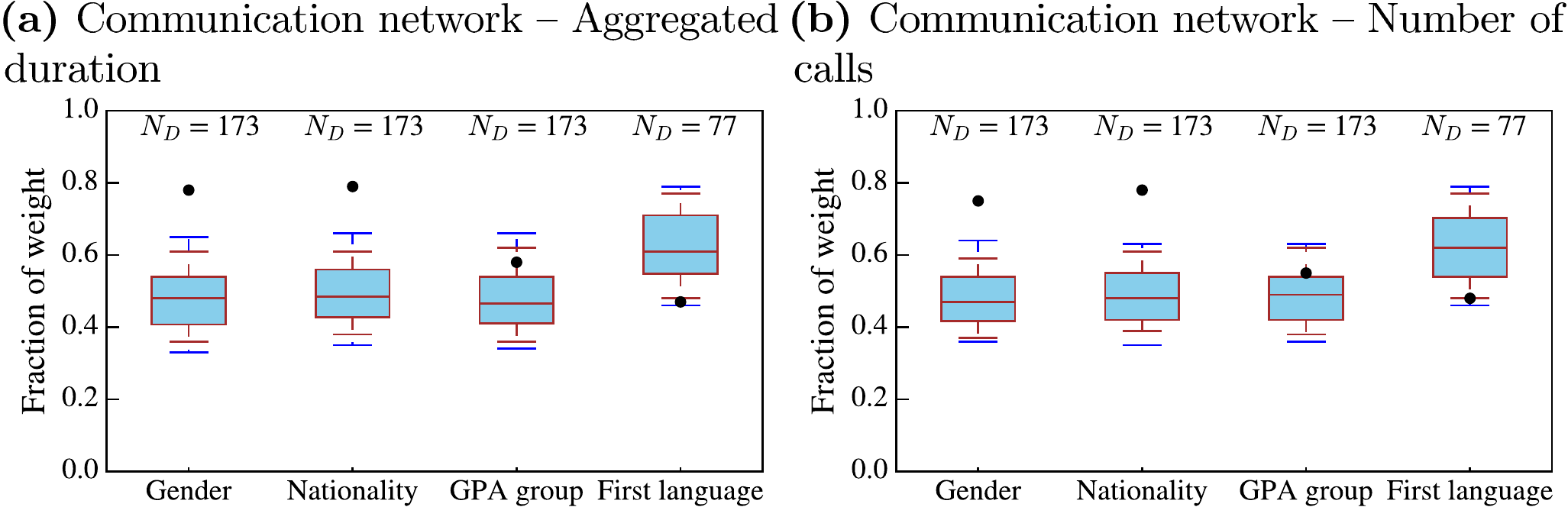}
\caption{{\bf Dyadic homophily -- Yearly-aggregated communication network.} Data (black dots) are compared with the distribution (boxplots) obtained for a null model in which attributes are reshuffled among nodes.
$N_D$ gives the number of dyads on which the measure is performed. 
Boxplot: Whiskers - $5^\th$, $10^\th$ and 
$90^\th$, $95^\th$ percentiles, Box - $25^\th$ and $75^\th$ percentiles.
} 
\label{fig5}
\end{figure}
\begin{figure}[h!]
  \includegraphics[width=0.9\textwidth]{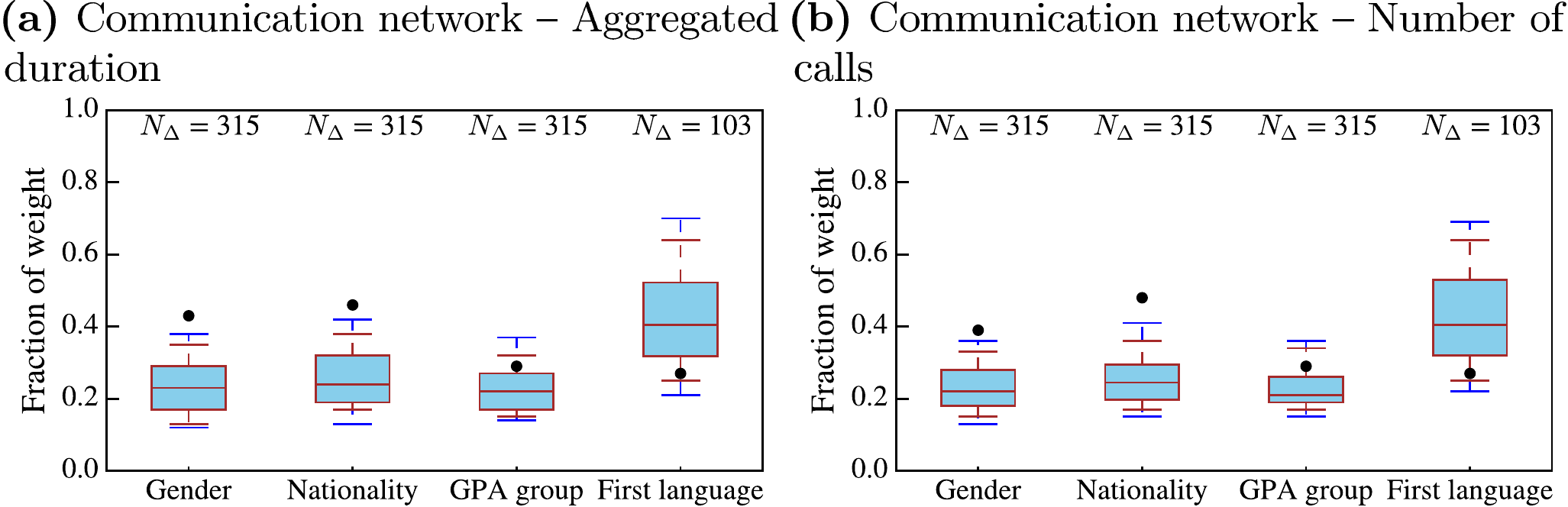}
\caption{{\bf Triadic homophily -- Yearly-aggregated communication network.}
Data (black dots) are compared with the distribution (boxplots) obtained for a null model in which attributes are reshuffled among nodes. $N_\Delta$ gives the number of triads on which the statistics is made.
Boxplot: Whiskers - $5^\th$, $10^\th$ and 
$90^\th$, $95^\th$ percentiles, Box - $25^\th$ and $75^\th$ percentiles.}
\label{figtottriad}
\end{figure}

Figures~\ref{fig5} and~\ref{figtottriad} show the homophily patterns with respect to gender, nationality, first spoken language and GPA uncovered by 
investigating the fraction of weight carried respectively by links and triangles between individuals with the same attribute, as described
in the Methods section. Very strong homophily patterns are found with respect to gender and nationality, not only at the dyadic level
but also for triangles: gender and nationality homophily determine which triangles, and not only which links, 
carry more weight in the network. Homophily with respect to GPA is on the other hand absent or at most very weak, while heterophilic patterns are observed for the first language.

\begin{figure}[h!]
  \includegraphics[width=0.9\textwidth]{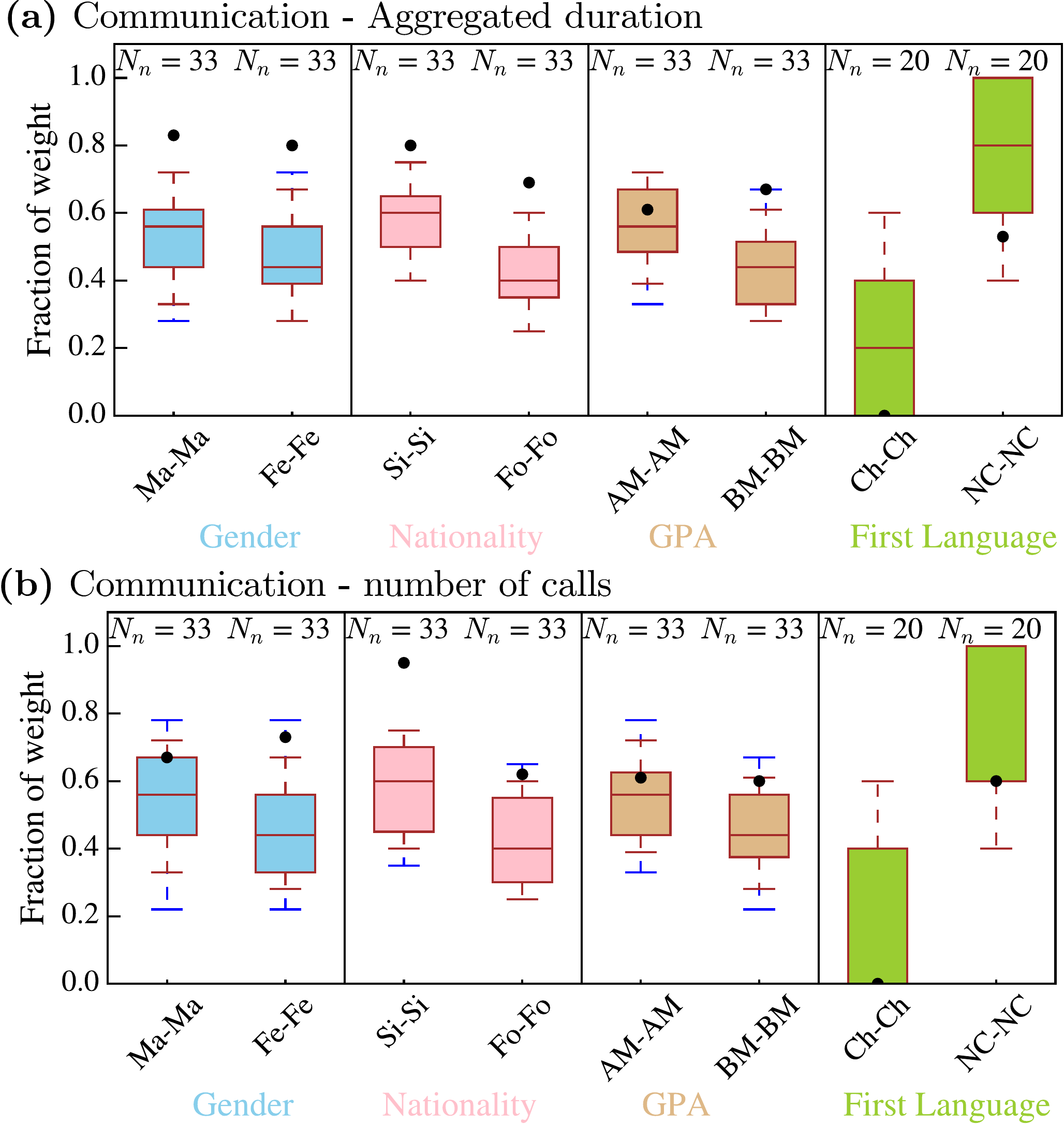}
\caption{{\bf Homophily in social preference -- Yearly-aggregated communication network} Ma: Male, Fe: Female, Si: Singaporean, Fo: Foreigner, AM: Above Median, BM: Below Median, Ch: Chinese, NC: Non-Chinese ; 
Data (black dots) are compared with the distribution (boxplots) obtained for a null model in which attributes are reshuffled among nodes. 
Boxplot: Whiskers - $5^\th$, $10^\th$ and 
$90^\th$, $95^\th$ percentiles, Box - $25^\th$ and $75^\th$ percentiles.}
\label{fig6}
\end{figure}

Figure~\ref{fig6} investigates the social preference homophily patterns of each group of individuals. Both male and female students show a clear
homophily pattern in their preferred communication partner. Similarly, both Singaporean and Foreigners display homophilous social preference.
On the other hand, homophily with respect to GPA shows contrasting trends: individuals with an above median GPA do not show homophily in their preferred communication partner, while individuals who have low GPA (below median) do (more so in terms of aggregated duration of communication than in terms of number of calls). For 
first spoken language, a weak tendency toward heterophily is observed for
non-chinese speaking students.

\begin{figure}[h!]
  \includegraphics[width=0.9\textwidth]{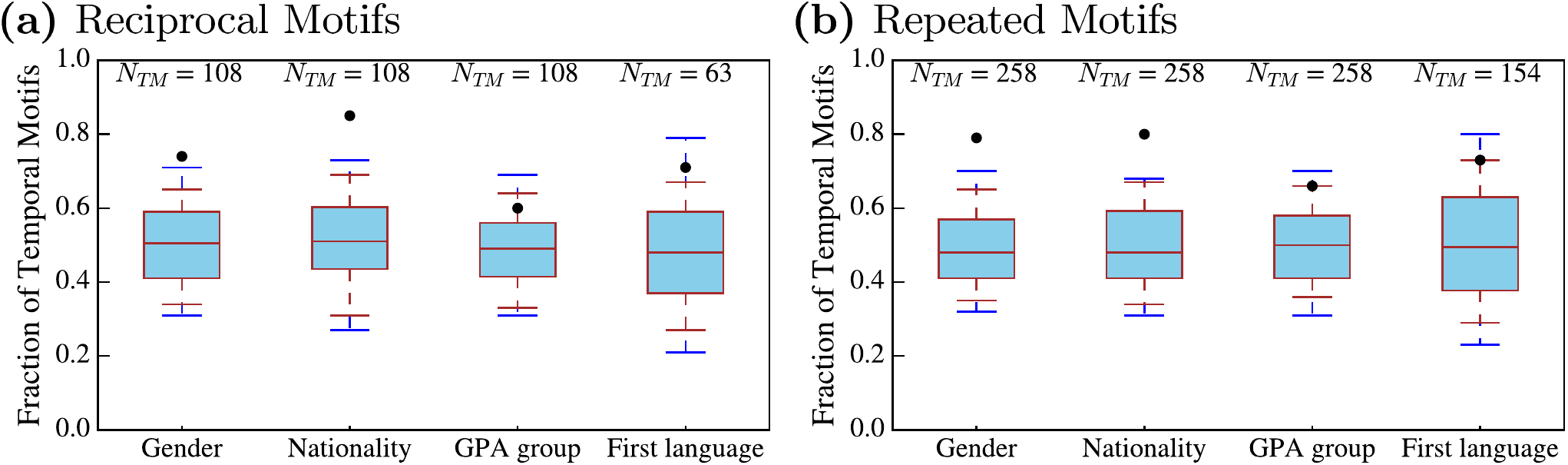}
\caption{{\bf Homophily in temporal motifs -- Yearly-aggregated communication network. Motif Types - Reciprocal and Repeated events}
Data (black dots) are compared with the distribution (boxplots) obtained for a null model in which attributes are reshuffled among nodes. 
$N_{TM}$ gives the number of temporal motifs in the network.
Boxplot: Whiskers - $5^\th$, $10^\th$ and 
$90^\th$, $95^\th$ percentiles, Box - $25^\th$ and $75^\th$ percentiles.
}
\label{fig7}
\end{figure}

Finally, Fig.~\ref{fig7} exhibits strong homophily patterns observed in reciprocal and repeated call motifs, both for gender and nationality. 
Only weak homophily is further observed with respect to GPA. In the first spoken language case, we also observe some tendency toward homophily, in contrast with the other indexes described above.

With respect to these attributes, various
homophily patterns are thus observed when aggregating over the whole dataset of one year 
without taking into account the timing of communication events, but also when considering sequences of calls separated by short time windows.

\subsection*{Evolution of homophily in communication across terms}

We now turn to the study of how homophily patterns evolve across the year in the group of students. To this aim, since questionnaire networks were collected once in each term, and also to work with sufficient statistics, we consider term-aggregated networks of communication. We show here the results
corresponding to homophily patterns in dyads, while 
figures for triadic homophily and social preference are
shown in the Supporting Information.
\begin{figure}[h!]
  \includegraphics[width=0.9\textwidth]{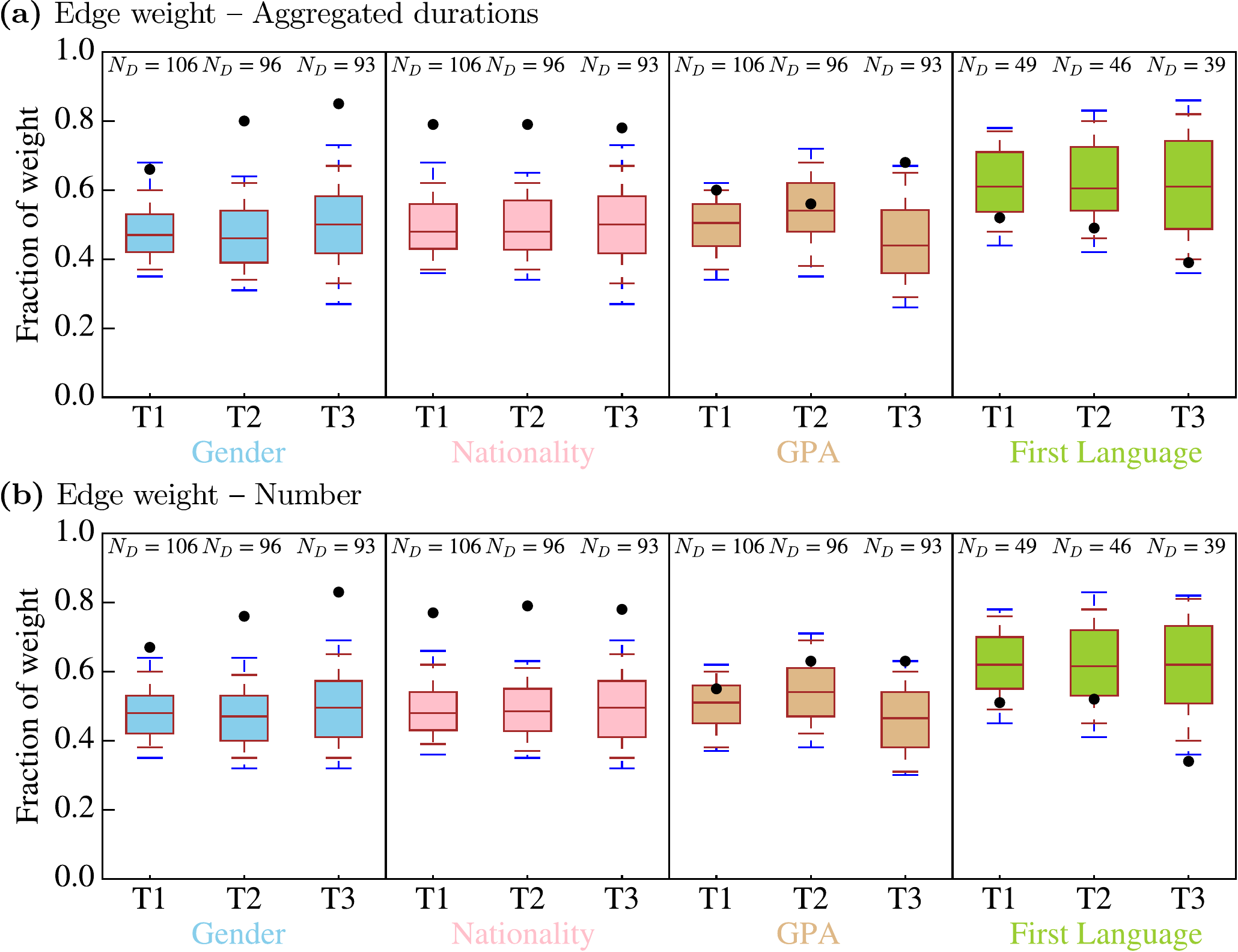}
\caption{{\bf Homophily in Dyads with respect to 
several attributes - Term aggregated communication networks.}
Data (black dots) are compared with the distribution (boxplots) obtained for a null model in which attributes are reshuffled among nodes. 
$N_D$ gives the number of dyads on which the measure is performed. 
Boxplot: Whiskers - $5^\th$, $10^\th$ and 
$90^\th$, $95^\th$ percentiles, Box - $25^\th$ and $75^\th$ percentiles.
}
\label{fig_evol}
\end{figure}
Gender homophily as revealed by the weight carried by dyads with the same gender is very strong in all terms, and exhibits a clear increasing trend (Fig. \ref{fig_evol}). The same increasing trend is observed in the weight carried by homophilic triads, even if the evidence for homophily is only weak with respect to the null model in the first term. In terms of social preference patterns,
 homophily increases for males, from absent or weak in the first two terms to very strong in the last term, while it is very strong in all terms for females (see Supporting Information). 
 
Homophily with respect to nationality is also very strong and stable across terms as measured by dyads. It weakens, however, in the third term as measured by triads. In terms of social preference, interesting distinct patterns are found: homophily decreases strongly and becomes weak or absent
in the third term for Singaporean students, but instead remain very strong and in fact increase for foreigners 
(see Supporting Information).

The tendency toward homophily with respect to GPA remains rather weak across all terms with respect to all indicators, except in the first term for triads and in the third term for dyads.
On the other hand, 
several instances of heterophilic tendencies are found
with respect to the first spoken language.
\begin{figure}[h!]
  \includegraphics[width=0.9\textwidth]{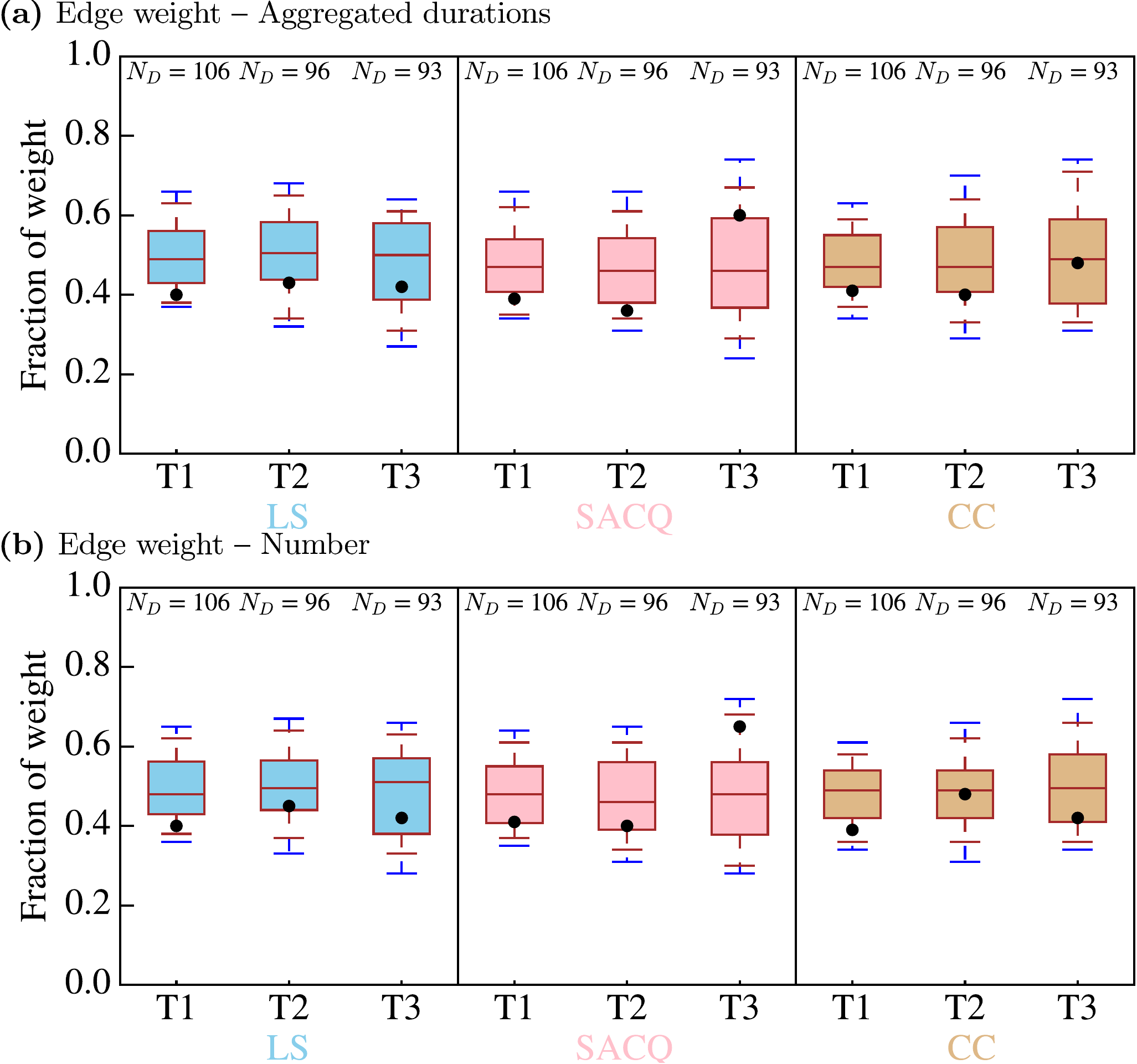}
\caption{{\bf Dyadic homophily with respect to 
psychological indices -- Term-aggregated communication networks.}
LS: loneliness scale; SACQ: student adaptation to college;
CC: classroom community scale (black dots) are compared with the distribution (boxplots) obtained for a null model in which attributes are reshuffled among nodes. 
$N_D$ gives the number of dyads on which the measure is performed. 
Boxplot: Whiskers - $5^\th$, $10^\th$ and 
$90^\th$, $95^\th$ percentiles, Box - $25^\th$ and $75^\th$ percentiles.}
\label{fig_evol_psy}
\end{figure}

Finally, we find no clear tendency toward homophilous behavior of students with respect to their scores in the three psychological questionnaires
(Fig.~\ref{fig_evol_psy}).
Some tendency toward heterophilous behavior is even observed in some cases, in particular in the social preference of the students with loneliness index below median.
\begin{table}[h!]
\caption{\bf Summary of the dyadic homophily patterns found in the different networks, with respect to the various attributes
considered.}
\small 
\centering 
\begin{tabular}{|p{1.8cm}|p{1cm}|p{0.7cm}|p{0.7cm}|p{0.75cm}||p{0.7cm}|p{0.7cm}|p{0.7cm}||p{0.8cm}|p{0.7cm}|p{0.7cm}|p{0.7cm}|p{0.7cm}|}
\hline
\multirow{2}{*}{Attribute} & \multirow{2}{*}{Weight} & \multicolumn{3}{|l||}{\parbox{1cm}{Communication}}  & \multicolumn{3}{|l||}{\parbox{2cm}{Co-presence}} & \multirow{2}{*}{Weight} & \multicolumn{4}{|l|}{ \parbox{1cm}{Questionnaires}} \\
\cline{3-8}
\cline{10-13}
&&T1 & T2 & T3 & T1 & T2 & T3 && T0&T1 & T2 & T3  \\
\hline
\multirow{2}{*}{Gender}
&Volume& \cellcolor{green!40} S& \cellcolor{green!90}VS&\cellcolor{green!90} VS& \cellcolor{green!40}S&\cellcolor{red!10} No&\cellcolor{green!90} VS& Q1&\cellcolor{green!90} VS&\cellcolor{green!40} S&\cellcolor{green!40} S&\cellcolor{green!90} VS \\
\cline{2-13}
&Number&\cellcolor{green!90} VS& \cellcolor{green!90}VS&\cellcolor{green!90} VS& \cellcolor{green!10}W&\cellcolor{red!10} No&\cellcolor{green!40} S& Q2&\cellcolor{green!40} S &\cellcolor{green!90}VS&\cellcolor{green!90} VS& \cellcolor{green!90}VS \\
\hline
\multirow{2}{*}{Nationality}
&Volume& \cellcolor{green!90}VS&\cellcolor{green!90} VS&\cellcolor{green!90} VS&\cellcolor{green!90} VS&\cellcolor{red!10} No&\cellcolor{red!10} No& Q1&\cellcolor{green!90} VS&\cellcolor{green!90} VS&\cellcolor{green!90} VS& \cellcolor{green!90} VS \\
\cline{2-13}
&Number&\cellcolor{green!90} VS&\cellcolor{green!90} VS&\cellcolor{green!90} VS& \cellcolor{green!90}VS&\cellcolor{red!10}No&\cellcolor{red!10}No& Q2&\cellcolor{green!90} VS& \cellcolor{green!10}W&\cellcolor{green!40} S&\cellcolor{green!10} W\\
\hline
\multirow{2}{*}{GPA}
&Volume&\cellcolor{green!10}W&\cellcolor{red!10} No&\cellcolor{green!90} VS&\cellcolor{green!90} VS&\cellcolor{green!90} VS&\cellcolor{green!90} VS& Q1& &\cellcolor{green!90}VS&\cellcolor{green!10} W&\cellcolor{green!10} W\\
\cline{2-13}
&Number&\cellcolor{red!10}No&\cellcolor{green!10} W&\cellcolor{green!40} S&\cellcolor{green!90} VS&\cellcolor{green!90} VS&\cellcolor{green!10} W& Q2&&\cellcolor{green!90} VS&\cellcolor{red!10} No&\cellcolor{green!10} W\\
\hline
\multirow{2}{*}{\parbox{1.5cm}{First Language}}
&Volume&\cellcolor{red!35}W$_\mathrm{het}$ &\cellcolor{red!35} W$_\mathrm{het}$ &\cellcolor{red!65}  S$_\mathrm{het}$ &\cellcolor{red!10}No &\cellcolor{red!10}No&\cellcolor{red!10} No& Q1&\cellcolor{red!10} No&\cellcolor{red!10} No&\cellcolor{red!10} No&\cellcolor{red!10} No\\ 
\cline{2-13}
\cline{2-13}
&Number&\cellcolor{red!35} W$_\mathrm{het}$ &\cellcolor{red!35} W$_\mathrm{het}$ &\cellcolor{red!95} VS$_\mathrm{het}$ &\cellcolor{red!10} No&\cellcolor{red!10} No&\cellcolor{red!10} No& Q2&\cellcolor{red!10} No&\cellcolor{red!10} No&\cellcolor{red!35} W$_\mathrm{het}$ &\cellcolor{red!10} No\\
\hline
\multirow{2}{*}{Loneliness}
&Volume&\cellcolor{red!35} W$_\mathrm{het}$ &\cellcolor{red!35} W$_\mathrm{het}$ &\cellcolor{red!10} No &\cellcolor{red!65} S$_\mathrm{het}$  &\cellcolor{red!10} No&\cellcolor{red!10} No& Q1&\cellcolor{green!90} VS&\cellcolor{red!10} No&\cellcolor{red!10} No&\cellcolor{red!10} No\\
\cline{2-13}
&Number&\cellcolor{red!35} W$_\mathrm{het}$ &\cellcolor{red!10} No&\cellcolor{red!10} No&\cellcolor{red!35} W$_\mathrm{het}$ &\cellcolor{red!10} No&\cellcolor{red!10} No& Q2&\cellcolor{red!35} W$_\mathrm{het}$ & \cellcolor{red!10} No &\cellcolor{red!10} No&\cellcolor{red!10} No\\
\hline
\multirow{2}{*}{\parbox{1.7cm}{SACQ}}
&Volume&\cellcolor{red!35} W$_\mathrm{het}$ &\cellcolor{red!25} W$_\mathrm{het}$ &\cellcolor{green!10} W& \cellcolor{green!10}W &\cellcolor{red!65} S$_\mathrm{het}$ &\cellcolor{red!35} W$_\mathrm{het}$ & Q1&\cellcolor{green!90} VS &\cellcolor{red!35} W$_\mathrm{het}$ &\cellcolor{red!10} No &\cellcolor{red!10} No\\ [6pt]
\cline{2-13}
&Number&\cellcolor{red!10}No  &\cellcolor{red!10} No&\cellcolor{green!10} W&\cellcolor{green!40} S&\cellcolor{red!35} W$_\mathrm{het}$ &\cellcolor{red!35} W$_\mathrm{het}$ & Q2&\cellcolor{green!10} W& \cellcolor{red!10}No&\cellcolor{red!10} No&\cellcolor{red!10} No\\ [6pt]
\hline
\multirow{2}{*}{\parbox{1.7cm}{Classroom Community}}
&Volume&\cellcolor{red!35} W$_\mathrm{het}$ & \cellcolor{red!35} W$_\mathrm{het}$ &\cellcolor{red!10} No&\cellcolor{green!10} W& \cellcolor{red!10} No&\cellcolor{red!10} No& Q1&\cellcolor{red!10} No&\cellcolor{red!10} No&\cellcolor{red!10} No&\cellcolor{green!10} W\\
\cline{2-13}
&Number&\cellcolor{red!35} W$_\mathrm{het}$&\cellcolor{red!10} No&\cellcolor{red!10} No&\cellcolor{green!10} W&\cellcolor{green!10} W&\cellcolor{red!10} No& Q2&\cellcolor{red!35} W$_\mathrm{het}$ &\cellcolor{green!10} W&\cellcolor{red!10} No&\cellcolor{red!10}No\\ 
\hline
\end{tabular}
\label{table:summary_dyads}
\end{table}

\subsection*{Comparison between homophily in various networks}

As discussed in the introduction, an important issue, besides the evidence for homophily (or the lack thereof) in each layer of interaction or relations
available for analysis, is whether the same or different conclusions are reached when investigating these different layers.
As made clear from the comparison reported above, there are indeed significant correlations between communication and
friendship or trust networks, and the students linked in the communication network tend also to have spent more time in co-presence. However, these networks are very distinct both in terms of structure and weights. 

\begin{figure}[h!]
  \includegraphics[width=0.9\textwidth]{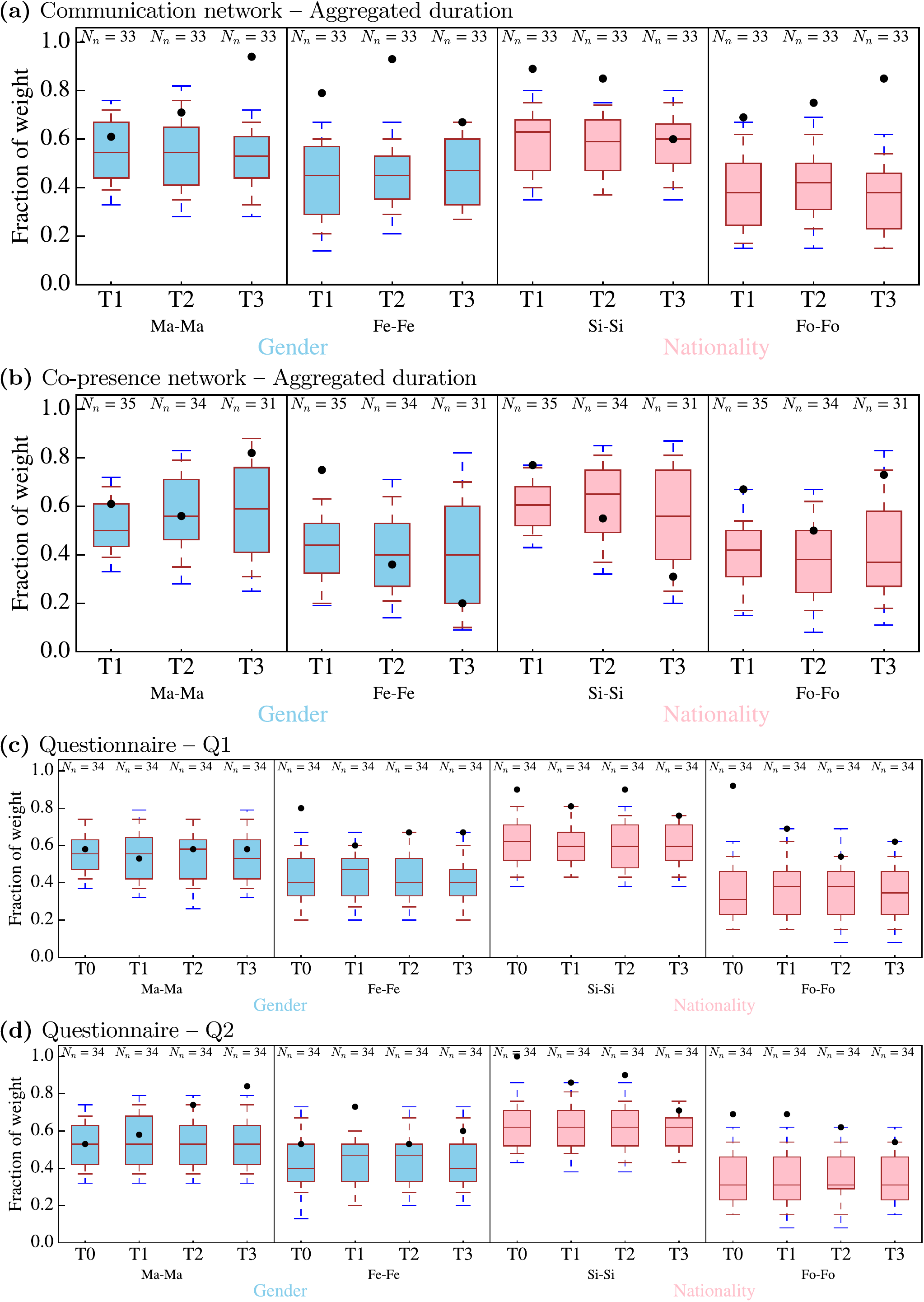}
\caption{{\bf Concordant vs. Discordant cases: Homophily in Social Preference in 
Gender and Nationality -- Term-aggregated networks.} Ma: Male, Fe: Female, Si: Singaporean, Fo: Foreigner;
Data (black dots) are compared with the distribution (boxplots) obtained for a null model in which attributes are reshuffled among nodes. 
Boxplot: Whiskers - $5^\th$, $10^\th$ and 
$90^\th$, $95^\th$ percentiles, Box - $25^\th$ and $75^\th$ percentiles.
}
\label{figcdsp}
\end{figure}

In order to investigate if the layers are similar enough in terms of the homophily patterns they exhibit, it is possible to thoroughly compare the results
provided in the previous section for the communication network and in the Supporting Information for other networks. 
For instance, a direct visual investigation can be performed
through figures such as Fig. \ref{figcdsp}.
A systematic side-by-side comparison for all pairs of layers and all possible indicators of homophily would however
be difficult and tedious to carry out. We therefore
propose the following methodology:
For each network, we build tables summarizing the evidence for homophily or heterophily (such as Table~\ref{table:summary_dyads}, see also Supporting Information)
and, for each pair of networks, we count 
the number of cases in which one network gives a certain answer while the other network gives another answer. We tabulate
these numbers for each pair of networks and show the full tables in the Supporting Information. In Table~\ref{table:comparison}, we show the outcome of 
a simplified counting procedure in which we group ``No'', ``W'' 
and ``W$_\mathrm{het}$'' as evidence for ``No homophily nor heterophily pattern'' 
 on the one hand and ``S'' and ``VS''
(resp. ``S$_\mathrm{het}$'' and ``VS$_\mathrm{het}$'')
as evidence for homophily (resp. heterophily) on the other hand. 
Note that this methodology could easily be adapted to answer more detailed comparisons, for instance by separating attributes into
different groups (e.g., considering only homophily with respect to psychological indices).

A first assessment of the results gathered in Table~\ref{table:comparison} indicates that concordant cases 
(on the diagonals) 
are far more numerous than discordant ones. It is, however, important to deepen our analysis as this overall observation might simply be due to the
large number of indicators showing an absence of homophilous patterns. Indeed, if we consider a large number of attributes and a large number of indicators, and
only few of them show evidence for homophily, then many concordant cases will be automatically observed, even if the few cases of homophily
are very different in distinct network layers. To check if this is indeed the case, we resort to a comparison with the following null model:
for each layer and each homophily indicator (dyadic, triadic or social preference), we reshuffle at random the answers
(``VS'', ``S'', ``W'', ``No'', ``W$_\mathrm{het}$'', ``S$_\mathrm{het}$'' and ``VS$_\mathrm{het}$'') across terms and attributes, and compute for each reshuffling the number of concordant and discordant
cases. We present in Table~\ref{table:comparison} the confidence intervals (C.I.) defined by the $5^\th$ and $95^\th$ percentiles of this null model, and we emphasize in boldface the cases in which
the empirical numbers are outside the C.I.

For the comparison between the two questionnaire networks, as well as between the communication network and the questionnaire networks, 
the numbers of concordant cases with and without homophily are both 
 much larger than the upper bound of the confidence intervals of the
 null model, while the numbers of cases in which one network
 shows homophily while the other does not are smaller than the  lower bound of the C.I. These three networks have therefore overall similar homophily patterns, despite discrepancies 
 occurring in a number of specific cases.

On the other hand, comparisons involving the co-presence network
lead mostly to numbers of concordant and discordant cases within the C.I. of the null model.  This means that, even if the 
co-presence network displays a similar ``amount'' of evidence for homophilous behavior with respect to the other layers of the social network, the homophily patterns are no more similar than random, given this amount. Hence, the co-presence homophily patterns do not inform us about which specific attributes and which specific indicators exhibit homophily patterns in the other networks.


\begin{table}[h!]
\caption{{\bf Comparison of homophily in networks.}
Each table corresponds to a pair of networks
and gives at row X and column Y the
number of cases in which an indicator gives a 
result X in the first network and a result Y in the second.
The intervals correspond to the confidence intervals
of the null model described in the text, and empirical numbers
are emphasized in boldface if they lie outside this interval.
}
\tiny
\begin{subtable}{1\textwidth}
\caption{\bf Communication vs Co-presence}
\centering 
\begin{tabular}
{|p{0.1cm}|p{0.7cm}|p{0.8cm}|p{0.8cm}|p{0.8cm}|}
\cline{3-5}
\multicolumn{1}{c}{}&\multicolumn{1}{c}{} & \multicolumn{3}{|c|}{Co-presence} \\ [1.5ex]
\cline{3-5}
\multicolumn{1}{c}{}&\multicolumn{1}{c|}{}&S/VS&No/W/ W$_\mathrm{het}$&S$_\mathrm{het}$/ VS$_\mathrm{het}$ \\ [1ex]
\hline
\multirow{3}{*}{\rotatebox[origin=c]{90}{Communication}} & 
S/VS& {\bf 16} [5-13]& {30} [30-38]& {\bf 0} [1-5] \\ [3ex]
\cline{2-5}
&No/W/ W$_\mathrm{het}$ & {\bf 17} [18-26]& 89 [83-91]& {\bf 11} [6-10] \\ [3ex]
\cline{2-5}
&S$_\mathrm{het}$/ VS$_\mathrm{het}$& 0 [0-3]& 5 [2-6]& 0 [0-1] \\[3ex]
\hline
\end{tabular}
\end{subtable}
  \hfill 
\begin{subtable}{1\textwidth}
\caption{\bf Questionnaire - Q1 vs Q2}
\centering 
\begin{tabular}
{|p{0.1cm}|p{0.7cm}|p{0.8cm}|p{0.8cm}|p{0.8cm}|}
\cline{3-5}
\multicolumn{1}{c}{}&\multicolumn{1}{c}{} & \multicolumn{3}{|c|}{Q2} \\ [1.5ex]
\cline{3-5}
\multicolumn{1}{c}{}&\multicolumn{1}{c|}{}&S/VS&No/W/ W$_\mathrm{het}$&S$_\mathrm{het}$/ VS$_\mathrm{het}$ \\ [1ex]
\hline
\multirow{3}{*}{\rotatebox[origin=c]{90}{Q1}} & 
S/VS& {\bf 18} [4-10]& {\bf 14} [19-26]& 0 [0-4] \\ [3ex]
\cline{2-5}
&No/W/ W$_\mathrm{het}$ & {\bf 18} [24-30]& {\bf 102}
[90-97]& 8 [6-9] \\ [3ex]
\cline{2-5}
&S$_\mathrm{het}$/ VS$_\mathrm{het}$& 0 [0-3]& 6 [4-8]& 2 [0-2] \\[3ex]
\hline
\end{tabular}
\end{subtable}
 \hfill 
\begin{subtable}{1\textwidth}
\caption{\bf Communication vs Q1}
\centering 
\begin{tabular}
{|p{0.1cm}|p{0.7cm}|p{0.8cm}|p{0.8cm}|p{0.8cm}|}
\cline{3-5}
\multicolumn{1}{c}{}&\multicolumn{1}{c}{} & \multicolumn{3}{|c|}{Q1} \\ [1.5ex]
\cline{3-5}
\multicolumn{1}{c}{}&\multicolumn{1}{c|}{}&S/VS&No/W/ W$_\mathrm{het}$&S$_\mathrm{het}$/ VS$_\mathrm{het}$ \\ [1ex]
\hline
\multirow{3}{*}{\rotatebox[origin=c]{90}{Communication}} & 
S/VS& {\bf 23} [6-12]& {\bf 22} [31-38]& 1 [1-4] \\ [3ex]
\cline{2-5}
&No/W/ W$_\mathrm{het}$ & {\bf 9} [19-25]& {\bf 102} [85-93]& 6 [3-7] \\ [3ex]
\cline{2-5}
&S$_\mathrm{het}$/ VS$_\mathrm{het}$& 0 [0-2]& 4 [3-6]& 1 [0-1] \\[3ex]
\hline
\end{tabular}
\end{subtable}
 \hfill 
\begin{subtable}{1\textwidth}
\caption{\bf Communication vs Q2}
\centering 
\begin{tabular}
{|p{0.1cm}|p{0.7cm}|p{0.8cm}|p{0.8cm}|p{0.8cm}|}
\cline{3-5}
\multicolumn{1}{c}{}&\multicolumn{1}{c}{} & \multicolumn{3}{|c|}{Q2} \\ [1.5ex]
\cline{3-5}
\multicolumn{1}{c}{}&\multicolumn{1}{c|}{}&S/VS&No/W/ W$_\mathrm{het}$&S$_\mathrm{het}$/ VS$_\mathrm{het}$ \\ [1ex]
\hline
\multirow{3}{*}{\rotatebox[origin=c]{90}{Communication}} & 
S/VS& {\bf 26} [6-12]& {\bf 20} [31-37]& {\bf 0} [1-5] \\ [3ex]
\cline{2-5}
&No/W/ W$_\mathrm{het}$ &{\bf 10} [22-29]& {\bf 97} [81-88]& {\bf 10} [5-9] \\ [3ex]
\cline{2-5}
&S$_\mathrm{het}$/ VS$_\mathrm{het}$& 0 [0-3]& 5 [2-6]& 0 [0-1] \\[3ex]
\hline
\end{tabular}
\end{subtable}
 \hfill 
\begin{subtable}{1\textwidth}
\caption{\bf Co-presence vs Q1}
\centering 
\begin{tabular}
{|p{0.1cm}|p{0.7cm}|p{0.8cm}|p{0.8cm}|p{0.8cm}|}
\cline{3-5}
\multicolumn{1}{c}{}&\multicolumn{1}{c}{} & \multicolumn{3}{|c|}{Q1} \\ [1.5ex]
\cline{3-5}
\multicolumn{1}{c}{}&\multicolumn{1}{c|}{}&S/VS&No/W/ W$_\mathrm{het}$&S$_\mathrm{het}$/ VS$_\mathrm{het}$ \\ [1ex]
\hline
\multirow{3}{*}{\rotatebox[origin=c]{90}{Co-presence}} & 
S/VS& 9 [3-10]& 24 [21-28]& 0 [0-4] \\ [3ex]
\cline{2-5}
&No/W/ W$_\mathrm{het}$ &23 [20-27]& 93 [91-99]& 8 [4-8] \\ [3ex]
\cline{2-5}
&S$_\mathrm{het}$/ VS$_\mathrm{het}$& 0 [0-4]& 11 [6-12]& 0 [0-2] \\[3ex]
\hline
\end{tabular}
\end{subtable}
 \hfill
\begin{subtable}{1\textwidth}
\caption{\bf Co-presence vs Q2}
\centering 
\begin{tabular}
{|p{0.1cm}|p{0.7cm}|p{0.8cm}|p{0.8cm}|p{0.8cm}|}
\cline{3-5}
\multicolumn{1}{c}{}&\multicolumn{1}{c}{} & \multicolumn{3}{|c|}{Q2} \\ [1.5ex]
\cline{3-5}
\multicolumn{1}{c}{}&\multicolumn{1}{c|}{}&S/VS&No/W/ W$_\mathrm{het}$&S$_\mathrm{het}$/ VS$_\mathrm{het}$ \\ [1ex]
\hline
\multirow{3}{*}{\rotatebox[origin=c]{90}{Co-presence}} & 
S/VS& {\bf 13} [4-11]& {\bf 18} [20-28]& 2 [0-4] \\ [3ex]
\cline{2-5}
&No/W/ W$_\mathrm{het}$ &23 [23-30]& 93 [86-95]& 8 [5-9] \\ [3ex]
\cline{2-5}
&S$_\mathrm{het}$/ VS$_\mathrm{het}$& 0 [0-5]& 11 [5-12]& 0 [0-2]\\[3ex]
\hline
\end{tabular}
\end{subtable}
 \hfill
\label{table:comparison}
\end{table}

\section*{Discussion}

The increased availability of data providing proxies for human behavior and social relationships, often in digital form, has led to a surge in the
number of studies of social theories and effects. Most such studies are, however, based on the analysis of one specific layer (e.g., phone call communications) of
the population social network, which is best represented as a multilayer network. It is now well established that the various network layers bear some level of correlations but are far from being equivalent. However, it is still unclear to what extent one can infer general conclusions from the study of only one
layer. In this paper, we have considered this issue--with a particular focus on homophily patterns---through the lens of a dataset providing data on several layers of the same population, namely
a communication layer, a co-presence layer, and two questionnaires describing friendship and trust relationships. The population under scrutiny
is formed of first-year students in an Asian university. Notably, the diversity of students in the population allows us to investigate homophily patterns
along several dimensions: gender, nationality, first spoken language, GPA and psychological indices assessed by questionnaires. It is worth adding that most studies about homophily reported in the literature are concerned with populations having a homogeneous composition in terms of nationality and first language~\cite{Mollgaard:2016,Kossinets:2009,Aiello:2010,Stehle:2013,Palchykov:2012,Kovanen:2013,Jo:2014}.

In terms of direct comparison between networks, we found
no correlation between the weights of links in the co-presence and communication network, but significant correlations between 
communication or co-presence and questionnaires networks. We also found a clear
correlation between communication (number and call volume) 
and reported friendship strength, confirming results of other authors with other types of population~\cite{Hill:2003,Roberts:2009,Kossinets:2009,Palchykov:2012,Jo:2014,Saramaki:2014}. This latter point stands in stark contrast with the absence of correlation between the amount of co-presence and friendship strength.

The strongest uncovered evidence of homophily is with respect
to gender and nationality in several indicators and layers, while
weaker evidence concerns homophily with respect to academic performance as measured by the GPA. No homophily was found with respect to the first spoken language nor
psychological indices (similarly to~\cite{Mollgaard:2016}, even if for different indices).

Most importantly, we have put forward here a systematic way of comparing homophily
patterns with respect to a heterogeneous group of attributes in the different layers of a social network. This methodology is based on counting the numbers of concordant
and discordant indicators of homophily in each pair of networks. As a large number of concordances might simply be due to a scarcity of indicators showing
homophily, a crucial point 
is to compare these numbers with a null model in 
which the results of the indicators are reshuffled within each network and type of indicator. 
If the observed number of concordant (resp. discordant)
cases lies above (resp. below) 
the confidence interval of this null model, 
it means that both networks yield an overall 
concordant picture of the homophily
patterns in the studied social network, in a way that is not 
simply due to an overall lack of homophily.
On the other hand, if the observed number of concordant cases falls within the confidence interval of the null model, we can conclude that one cannot extract information about homophily patterns in one network from the patterns in the other network.

In the specific case under study, we found that the communication and questionnaire layers 
lead to similar conclusions in many cases--even if some minor discrepancies are observed--and more than expected from the null model. 
This means that the communication layer allows us to obtain information about homophilous trends in the friendship and trust networks of this social network. On the other hand, the co-presence network cannot be used to assess homophily patterns occurring in the other layers.

Our work has several limitations that are worth mentioning. Obviously, it is based on one single dataset of a specific population of limited size. The population was, however, largely isolated, and data is available with a high resolution in time for a whole year, allowing the analysis of the temporal evolution of the homophily patterns, as well as the comparison with the temporal evolution in the other layers.
Moreover, we could not reliably use messaging data, although messages nowadays represent a fair amount of communication between individuals. Furthermore, we did not have access to any online social network on which messages are also exchanged. The co-presence data had limited spatial resolution owing to the particular choice of the Bluetooth technology. It might be that with another technology yielding a higher spatial resolution, data on face-to-face interactions would lead to different conclusions, and correspond to a larger similarity of homophily patterns with the communication and questionnaire networks. 

To conclude, we note that the methodology put forward to assess the similarity of homophily patterns in different layers of a social network is general and can be
applied to any dataset composed of several layers of interactions or relationships between individuals, and to any set of attributes for which homophily patterns are of interest. 
We therefore hope that the present study will stimulate further similar dataset collections and investigations into this crucial issue.


\begin{backmatter}

\section*{Competing interests}
  The authors declare that they have no competing interests.

\section*{Author's contributions}
W.Q.Y. designed the network smartphone study and data collection.\\
A.B. and R.B. conceived and designed the study of homophily for the social network, with consultations from W.Q.Y.\\
A.M. performed the statistical data analysis. A.B., A.M., and R.B. wrote the manuscript. W.Q.Y. reviewed the manuscript.

\section*{Acknowledgements}
We would like to thank Ms. Xiaoqian Li for her assistance in making the database available to us, in answering multiple rounds of questions regarding the dataset, and in reviewing the manuscript.

This work was supported by the SUTD-MIT International Design Center (IDC) under Grant IDG31100106 and IDD41100104 (A.M., W.Q.Y., and R.B.).


\bibliographystyle{bmc-mathphys} 
\bibliography{bmc_article}      




\section*{Figures}















\section*{Tables}





%
\section*{Additional Files}
  \subsection*{Additional file 1 --- Supporting Information: Tables and Graphs}
{\bf Comparison between term-aggregated networks; Homophily patterns in the yearly aggregated co-presence network; Evolution of homophily patterns in triads
and social preference in the communication network;
Summary tables of homophily patterns for triads
and social preferences; Detailed tables of the numbers of concordant
and discordant cases for each network pair.} 

\end{backmatter}
\end{document}